\documentclass[journal]{IEEEtran}
\usepackage{amsmath,amsfonts,hyperref}
\usepackage{amsmath}
\usepackage{algorithmic}
\usepackage{array}
\usepackage[caption=false,font=normalsize,labelfont=sf,textfont=sf]{subfig}
\usepackage{textcomp}
\usepackage{stfloats}
\usepackage{url}
\usepackage{verbatim}
\usepackage{graphicx}
\usepackage{environ}
\usepackage{enumitem}
\usepackage{svg}
\usepackage{multirow} 
\usepackage{booktabs} 
\usepackage{newtxtext}
\usepackage{amsmath,amsfonts,hyperref}
\usepackage{amsthm}
\usepackage{orcidlink}
\usepackage{bm}
\usepackage{tikz}
\usepackage{xcolor}

\NewEnviron{added}[1]{%
  \BODY
}

\usepackage{microtype}
\usepackage{relsize} 
\usepackage{mathtools} 

\def\BibTeX{{\rm B\kern-.05em{\sc i\kern-.025em b}\kern-.08em
    T\kern-.1667em\lower.7ex\hbox{E}\kern-.125emX}}
\usepackage{balance}

\begin{document}

\title{\LARGE \bf
Two-Stage Bidirectional Inverter Equivalent Circuit Model for Distribution Grid Steady-State Analysis and Optimization
}
\author{
Emmanuel~O.~Badmus\,\orcidlink{0000-0003-4815-1708},~Graduate~Student~Member,~IEEE,
and~Amritanshu~Pandey\,\orcidlink{0000-0002-4431-6889},~Senior~Member,~IEEE

% \thanks{This material is based upon work supported by the U.S. Department of Energy’s award number DE-EE0010407 and award number DE-AC02-05CH11231. The views expressed herein do not necessarily represent the views of the U.S. Department of Energy or the United States Government.}% <-this % stops a space
\thanks{E. Badmus and A. Pandey are both with the Department of Electrical and Biomedical Engineering at the University of Vermont, Burlington, VT. {\tt\small {ebadmus, apandey1}@uvm.edu}}%
\thanks{Accepted for publication in IEEE Transactions on Power Systems.}
}

\markboth{IEEE TRANSACTIONS ON POWER SYSTEMS, VOL. XX, NO. X, XXX 2025}{}

\maketitle

\newcommand{\Inverter}{Two-Stage Bidirectional Inverter}
\newcommand{\InverterShort}{TSBI}
\newcommand{\FSC}{FSC}
\newcommand{\SSC}{SSC}
\newcommand{\TOne}{T1}
\newcommand{\TTwo}{T2}

\newcommand{\ForwardEnergyStorageMode}{Forward Energy Storage Mode}
\newcommand{\FESM}{FESM}
\newcommand{\ForwardEnergyTransferMode}{Forward Energy Transfer Mode}
\newcommand{\FETM}{FETM}
\newcommand{\ReverseEnergyStorageMode}{Reverse Energy Storage Mode}
\newcommand{\RESM}{RESM}
\newcommand{\ReverseEnergyTransferMode}{Reverse Energy Transfer Mode}
\newcommand{\RETM}{RETM}
\newtheorem{remark}{\bf Remark}

\begin{abstract}

% This paper presents an exact circuit-based steady-state model of a two-stage bidirectional inverter designed for low-voltage distributed energy resources, such as photovoltaic (PV) and battery systems.
% Unlike existing models that rely on constant efficiency assumptions, empirical loss data, or analytical loss approximations to represent inverter behavior and use mathematical abstractions to embed the mode of operation, the proposed model explicitly includes switching and conduction losses in both DC–DC and DC–AC converters. It uses continuous, twice-differentiable expressions that enable bidirectional power flow without binary variables or complementarity constraints, making it generic and applicable to all operating modes and device types (e.g., PV and battery systems).
% This paper presents an exact circuit-based steady-state model of a two-stage bidirectional inverter intended for photovoltaic (PV) and battery systems in residential distribution networks.
This paper presents a \textit{physics-based} steady-state equivalent circuit model of a two-stage bidirectional inverter.
These inverters connect distributed energy resources (DERs), such as photovoltaic (PV) and battery systems, to distribution grids.
Existing inverter models have technical gaps on three fronts:
i) inadequate modeling of inverter losses; %that are based on constant efficiency assumptions, empirical loss data, or analytical loss approximations, 
ii) use of mathematical abstractions for bidirectional flow of power; and
iii) inability to integrate different control modes into nonlinear solvers
%such as Volt-Var and constant power factor, 
without loss of generality.
We propose an inverter model that explicitly captures losses in passive circuit components, grounded in circuit-level principles.
% We enable bidirectional power flow without binary variables or complementarity constraints by incorporating ...
We enable bidirectional power flow without binary or complementarity constraints by formulating loss terms as smooth, sign-aware expressions of current.
% The proposed model supports the analysis of various active and reactive power control strategies, including maximum PowerPoint tracking (MPPT), constant power factor, and Volt–VAR regulation. 
We introduce and parameterize controlled current sources with twice-differentiable continuous functions to enable inverter control modes without loss of generality.
%The proposed model accommodates a range of active and reactive power control strategies—including maximum power point tracking (MPPT), constant power factor, and Volt–VAR regulation—making it suitable for analysis and optimization studies.
% Optimization results on large-scale distribution feeders show that incorporating nonlinear loss mechanisms and bidirectional inverter control captures efficiency trends and reactive power interactions more accurately than constant-efficiency or empirical models. 
% We validated the model against time-domain simulations in Simulink. We applied it to power flow analysis on a 16k+ node network, single-house energy dispatch optimization, and PV curtailment optimization on a realistic 20k+ node distribution system.
% We validated the proposed model against time-domain simulations in Simulink and integrated it at the load buses of distribution networks to perform power flow analysis on a 16k+ node network, single-house energy dispatch optimization, and PV curtailment optimization on a realistic 20k+ node network.
We integrate DERs with the proposed inverter model at the load buses of distribution networks to perform power flow and optimization studies on real-world distribution networks with over 20,000 nodes.
%perform power flow analysis on a 16k+ node network, single-house energy dispatch optimization, and PV curtailment optimization on a realistic 20k+ node network.
% Results highlight how reactive power modulation impacts voltage profiles, conversion efficiency, and battery state-of-charge estimates under varying generation and load conditions. 
We demonstrate that the proposed model is more accurate, integrates seamlessly with various control modes without loss of generality, and scales robustly to large optimization problems.
%more accurately captures i) inverter efficiency, ii) reactive power behavior, and iii) battery state-of-charge, while maintaining computational tractability and numerical stability when deployed across real-world large-scale distribution networks.
% By providing a unified, physics-based representation of internal inverter mechanics, the proposed TSBI model enables more accurate and calculations and grid optimization studies in high-penetration renewable environments, addressing key limitations of existing steady-state approaches. 


\end{abstract}

\begin{added}{Reviewer 2 Comment 6}
\begin{table}[htpb]
\centering
\renewcommand{\thetable}{}        % remove table number
\caption{List of Abbreviations}
\begin{tabular}{l p{6cm}}
\hline
\textbf{Abbrev.} & \textbf{Meaning / Description} \\
\hline
BESS & Battery Energy Storage System \\
CE-B & Constant-Efficiency Binary inverter model \\
CE-CS & Constant-Efficiency Complementarity Slackness inverter model \\
CPF & Constant Power Factor control mode \\
DER & Distributed Energy Resource \\
ECF & Equivalent Circuit Formulation \\
ECM & Equivalent Circuit Model \\
FSC & First-Stage Converter \\
HEMS & Home Energy Management System \\
KCL & Kirchhoff’s Current Law \\
KVL & Kirchhoff’s Voltage Law \\
LCL & Inductive–Capacitive–Inductive AC filter \\
MPPT & Maximum Power Point Tracking \\
PCC & Point of Common Coupling \\
PF & Power Factor \\
PV & Photovoltaic System \\
SSC & Second-Stage Converter \\
TSBI & Two-Stage Bidirectional Inverter \\
UPF & Unity Power Factor control mode \\
VV & Volt–VAR control mode \\
\hline
\end{tabular}
\label{tab:abbrev}
\end{table}
\addtocounter{table}{-1}   % prevent table counter increment

% ====== LIST OF SYMBOLS ======
\begin{table}[t]
\centering
\renewcommand{\thetable}{}        % remove table number
\caption{List of Symbols}
\begin{tabular}{l l l}
\hline
\textbf{Symbol} & \textbf{Meaning / Description} & \textbf{Units} \\
\hline
$C, C_f$ & Filter capacitance & F, µF \\
$C_{e,t}, C_{i,t}$ & Export/import electricity price at time $t$ & \$/kWh \\
$D$ & Duty cycle of FSC & -- \\
$E_{\text{cap}}$ & Battery energy capacity & kWh \\
$E_{\max}$ & Maximum battery energy capacity & kWh \\
$f_{\text{sw}}$ & Switching frequency & kHz \\
$f(x;a,b)$ & Smooth indicator for Volt–VAR shaping & -- \\
$I_0$ & PV diode saturation current & A \\
$I_{\text{AC}}^{R}, I_{\text{AC}}^{I}$ & Real/imaginary AC-side current components & A \\
$I_{\mathrm{AC}}, V_{\mathrm{AC}}$ & AC-side current and voltage magnitudes & A, V \\
$I_{\mathrm{DC}}, V_{\mathrm{DC}}$ & DC-link current and voltage & A, V \\
$I_{\mathrm{rms}}$ & RMS current magnitude & A \\
$I_{\text{ph}}$ & PV photocurrent & A \\
$I_{T2}^{R}, I_{T2}^{I}$ & AC side injection current components & A \\
$L_1, L_2$ & Converter- and grid-side inductances & H \\
$M$ & Modulation index of SSC & -- \\
$M^R, M^I$ & Real/imaginary components of $M$ & -- \\
$N_s$ & Number of PV cells in series & -- \\
$P_{\text{ctrl}}, Q_{\text{ctrl}}$ & Active/reactive control inputs & W / var \\
$P_{\mathrm{loss}}, P_{\mathrm{out}}$ & Total inverter loss / output power & W \\
$P_{\mathrm{loss,inv}}$ & Total inverter loss (FSC+SSC+filter) & W \\
$P_{\mathrm{MP}}$ & PV maximum power under MPPT & W \\
$P_{\text{set}}, Q_{\text{set}}$ & Constant-$P$ / constant-$Q$ setpoints & W / var \\
$P_{c,t}, P_{d,t}$ & Battery charge/discharge power at $t$ & W \\
$P_{batt,t}$ & Net battery power (positive = discharge) & W \\
$P_{\text{cu},n}$ & Curtailed active power for inverter $n$ & W \\
$P_{\text{exp},n}$, $P_{\text{exp},n}^{*}$ & Actual / desired export power & W \\
$Q_{\mathrm{RR}}$ & Diode reverse-recovery charge & C \\
$R_1, R_2, R_d$ & Filter and damping resistances & $\Omega$ \\
$R_L$ & Inductor winding resistance & $\Omega$ \\
$R_{\mathrm{int}}$ & Battery internal resistance & $\Omega$ \\
$R_{\mathrm{S}}, R_{\mathrm{SH}}$ & PV series / shunt resistances & $\Omega$ \\
$R_T, R_D$ & Transistor / diode on-state resistances & $\Omega$ \\
$S$ & Apparent power rating of inverter & kVA \\
$T_s$ & PWM switching period & s \\
$V_0$ & On-state voltage drop  & V \\
$V_D$ & $V_{\mathrm{MP}}+I_{\mathrm{MP}}R_s$ auxiliary PV term & V \\
$V_{D0}$ & Diode forward voltage drop & V \\
$V_{\mathrm{MP}}, I_{\mathrm{MP}}$ & PV MPP voltage and current & V, A \\
$V_{\mathrm{OC,nom}}$ & Nominal battery open-circuit voltage & V \\
$V_{\mathrm{th}}$ & Thermal voltage ($n N_s V_t$) & V \\
$V_t$ & Thermal voltage per cell ($kT/q$) & V \\
$V_{T0}$ & MOSFET threshold voltage & V \\
$V_{T2}^{R}, V_{T2}^{I}$ & PCC voltage components & V \\
$X_{L1}, X_{L2}, X_C$ & Reactances of filter elements & $\Omega$ \\
$\bm{\Psi}$ & Parameter set & mixed \\
$\chi$ & Exponential term $\exp(V_D/V_{\mathrm{th}})$ & -- \\
$\Delta \tau$ & Optimization time-step duration & h or s \\
$\epsilon$ & Smoothing constant $|x|_\epsilon=\sqrt{x^2+\epsilon}$ & -- \\
$\eta(P,Q)$ & Efficiency as a function of $P,Q$ & -- \\
$\eta_c, \eta_d$ & Fixed charge/discharge efficiencies & -- \\
$\eta_{\max}$ & Peak inverter efficiency & -- \\
$\phi$, $\varphi$ & Power-factor angle & rad \\
$\omega$ & Angular frequency ($2\pi f$) & rad/s \\
$\overline{I}_T,\overline{I}_D$ & Average transistor/diode current & A \\
$\varepsilon(x)$ & Percentage error $|x_{\mathrm{TSBI}}{-}x_{\mathrm{Sim}}|/|x_{\mathrm{Sim}}|$ & \% \\
$\|\cdot\|_{\infty}$ & Infinity-norm  & -- \\
\hline
\end{tabular}
\label{tab:symbols_all}
\end{table}
\addtocounter{table}{-1}   % prevent table counter increment
\end{added}

\section{INTRODUCTION}
\subsection{Research Motivation}

\noindent \IEEEPARstart{T}{he} increasing adoption of distributed energy resources (DERs), particularly photovoltaic (PV) systems and battery energy storage systems (BESS), is reshaping the operational behavior of distribution networks. 
These resources, interfaced through power electronic inverters, now influence local voltage profiles and alter the direction of power flow. 
Among available inverter topologies, hybrid grid-tied inverters that interface both PV and battery systems are becoming common in residential DER installations \cite{hasan2023performance}. These inverters typically employ a two-stage design, where a DC-DC converter adjusts the PV and battery voltages to a common DC link voltage, and a DC-AC converter supplies power to the grid. This same structure appears in multi-port versions, where each port uses its own DC--DC interface or a shared multi-input converter, all feeding a single DC link that is connected to the same DC--AC stage.

% Despite the growing importance of inverter-based DERs, most steady-state \InverterShort{} models in power flow and optimization studies do not accurately capture the underlying converter behavior. 
% Most models assume fixed efficiency or rely on simplified loss approximations, neglecting key nonlinear effects of conduction losses, switching transitions, and the operating control mode.
% As a result, they fail to capture critical phenomena, including lower efficiency under light loading, reactive power-related losses, and voltage-dependent performance variations. 
% Moreover, bidirectional operation is often modeled using binary variables or mode-specific constraints, leading to formulations that increase solver complexity and reduce scalability. 

Despite the growing importance of inverter-based DERs, most steady-state inverter models in power flow and optimization studies do not accurately capture the underlying converter behavior.
%Most models assume fixed efficiency or rely on simplified loss approximations \cite{jantsch1993results,braun2007reactive,bower2004performance,yen2010overall,schafmeister2005analytical}, neglecting key nonlinear effects of conduction losses, switching transitions, and the operating control mode. 
%As a result, they fail to capture critical phenomena, including lower efficiency under light loading, reactive power-related losses, and voltage-dependent behaviour. 
%Moreover, bidirectional power flow is often modeled using binary variables or complementary constraints \cite{elsaadany2023battery,leyffer2006complementarity}, introducing discrete or non-smooth formulations that lead to formulations that increase solver complexity and reduce scalability. 
%Also, some models apply control via outer-loop, decoupling control from inverter electrical behavior \cite{muthukaruppan2024supervisory} and exhibiting poor convergence due to the use of piecewise discontinuous models.
% The proposed \InverterShort{} model is intended for distribution utilities, grid operators, and researchers engaged in the analysis and optimization of inverter-dominated distribution systems. By providing a continuously differentiable, loss-aware representation of inverter physics and control behavior that is consistent with modern interconnection standards, such as IEEE 1547, the model enables scalable studies, such as hosting-capacity assessment, Volt–VAR coordination, and DER scheduling.
\begin{added}{Editor Comment 3 and Reviewer 3 Comment 1}
%The proposed %\InverterShort{} model is intended for distribution utilities, grid operators, and researchers engaged in the analysis and optimization of inverter-dominated distribution systems. 
We posit that a more accurate loss-aware representation of inverter physics, with control behavior consistent with modern interconnection standards such as IEEE 1547, will enhance the accuracy of key utility-run distribution grid analyses and optimizations, including dynamic hosting capacity, Volt–VAR coordination, and DER scheduling.
%we propose a model that is both accurate and scalable.
%and which can be used in studies that require scalability, such as 
\end{added}

% Also, most models implement control through outer-loop setpoints, which decouple the inverter’s electrical behavior from its control response \cite{xu2008parallel,jha2019bi}.

%In addition to energy conversion, inverter power control determines how DERs adjust their active and reactive outputs in response to network conditions. 
% Also, most current steady-state models of inverters treat control modes externally or reduce them to fixed setpoints, limiting their ability to reflect the inverter’s adaptive role in the networks. 

% We build a physics-based scalable steady-state model for \InverterShort{} that accurately models the internal losses, supports generic control modes, and enables seamless bidirectional operation across diverse loading and network conditions.

\subsection{State-of-the-Art Solutions and Limitations}

\noindent
With a focus on system-wide distribution grid analysis and optimization, current bidirectional inverter models exhibit challenges in three key areas:

% \noindent \textit{1) Inadequate loss representation:}
% Existing steady-state models model inverter losses with i) constant efficiency factors, ii) empirical data fits, or iii) simplified analytical expressions. 
% Constant efficiency models apply a fixed loss factor and disregard variations in efficiency resulting from load, voltage, or control strategy. 
% Empirical models, such as the Schmidt–Sauer \cite{jantsch1993results} and Braun \cite{braun2007reactive} formulations, fit quadratic functions to measured inverter performance data, typically as a function of active power output. 
% Standardized efficiency curves like those from the California Energy Commission (CEC) \cite{bower2004performance} and European weighted \cite{yen2010overall} benchmarks also fall into this category; they provide test-based, device-specific efficiency metrics under controlled conditions, often assuming unity power factor and fixed terminal voltage. 
% Analytical models, in contrast, derive losses from physics principles—using expressions based on switching frequency and semiconductor characteristics—to estimate conduction and switching losses \cite{schafmeister2005analytical}. 
% These models offer broader applicability, but all current models omit nonlinear dependencies on reactive power and voltage variations. 
% As a result, none of the existing approaches fully capture inverter loss behavior across the whole operating space.

\begin{added}{Editor Comment 1, Reviewer 1 Comment 1, and Reviewer 2 Comment 2}
\noindent \textit{1) Inadequate Loss Representation:}
Existing inverter models used in distribution-grid analysis and optimization represent conversion losses through three main abstractions: 
(i) constant-efficiency factors, 
(ii) standardized test-based metrics,
(iii) empirical data-fitting models, and
(iv) semi-analytical models.

Constant-efficiency formulations assume a fixed loss factor and disregard variations in efficiency due to load, voltage, or control strategy. 
In contrast, several studies show that inverter efficiency depends on loading \cite{allenspach2023power,anderson2022effect,thiagarajan2019effects}, terminal voltage \cite{gonzalez2014effect}, and reactive-power injection \cite{adak2025control,grab2022modeling,malamaki2019estimation}.
Standardized test-based metrics, including the CEC (California Energy Commission) \cite{bower2004performance} and European \cite{yen2010overall} weighted efficiencies, define load-weighting factors to yield reference efficiency values under controlled operating conditions. 
Although widely used for benchmarking, these metrics reflect only nominal test conditions and cannot generalize across varying operating voltages or control modes. 
Empirical data-fitting models, mostly derived from the Schmidt–Sauer \cite{jantsch1993results} and Braun \cite{braun2007reactive} formulations, provide continuous efficiency curves by fitting quadratic functions to measured inverter data. These models reproduce the general trend with output power but remain device-specific and insensitive to voltage or power-factor variations.
Semi-analytical models \cite{grab2022modeling,malamaki2019estimation} combine physics-based loss equations with calibrated coefficients to improve accuracy but still require parameter tuning for each inverter based on measured performance data.

\end{added}

% \noindent \textit{2) Challenges in bidirectional flow modeling:}
% Current techniques model bidirectional power flow in steady-state formulations through the introduction of two unidirectional branches-one for charging and one for discharging—along with mathematical constraints, such as binary variables or
% complementarity relations, to prevent simultaneous operation \cite{tant2012multiobjective}. 
% Binary-based models enforce operation exclusivity with integer variables, resulting in mixed-integer formulations that scale poorly as the number of inverters increases. 
% This is especially true when applied across time horizons in quasi-static time series optimization \cite{antic2024impact}. Complementarity constraints-based approaches, on the other hand, avoid binary variables but introduce non-differentiable and nonconvex structures, resulting in frequent local minima and saddle points  \cite{leyffer2006complementarity}. 

% Other forms of implementing this complementarity behavior include relaxed and penalty-based formulations  \cite{garifi2020convex}, continuous convex formulations  \cite{elsaadany2025linear, nazir2021guaranteeing, taha2024lossy} and convex energy envelopes.
% However, these methods simplify computation but can yield infeasible inverter states.

\begin{added}{ Editor Comment 1 and Reviewer 1 Comment 1}
\noindent \textit{2) Challenges in Bidirectional Flow Modeling:}
Current techniques model bidirectional power flow (for battery + inverter configurations) in steady-state formulations by introducing two unidirectional branches—one for charging and one for discharging—along with mathematical constraints to enforce complementarity (i.e., no simultaneous charge and discharge) \cite{tant2012multiobjective}. 
\cite{antic2024impact} enforces complementarity with binary variables. 
The resulting mixed-integer formulation scales poorly as the number of inverters increases. 
\cite{leyffer2006complementarity} use continuous complementarity constraints and avoid binary variables but introduce non-differentiable and nonconvex structures that often lead to local minima and saddle points. 
\cite{garifi2020convex} proposes relaxed and penalty-based formulations to improve tractability, and \cite{elsaadany2025linear} achieves realizable power schedules without explicit complementarity by using linear aggregate battery models.
Plus, there are works on convex relaxations of complementarity constraints \cite{nazir2021guaranteeing}, including convex energy-space representations \cite{taha2024lossy}. 
% However, these methods represent a single physical bidirectional flow using separate charge and discharge variables, highlighting the need for a single formulation that preserves physical consistency across operating modes.
Taken together, these approaches introduce either mixed integer complexity, highly non-convex constraints, or some form of relaxation. This gap can be addressed with a single variable formulation that captures both forward and reverse power flow through one variable.
\end{added}

\begin{added}{Editor Comment 1 and Reviewer 1 Comment 1}

\noindent \textit{3) Limited Integration of Control Logic:}  
\noindent Modern grid standards, including IEEE 1547-2018 \cite{photovoltaics2018ieee}, require inverters to provide reactive-power support and voltage regulation. 
Several steady-state models implement these functions using an outer loop around the power flow or optimization solvers with discontinuous piecewise control curves,  \cite{muthukaruppan2024supervisory,azzolini2021evaluating}, where inverter setpoints are updated between successive solves.
Such formulations often lead to numerical oscillations and poor convergence. 
To avoid iterative outer loops, \cite{jha2019bi} embeds control curves directly within optimization problems by representing piecewise-linear functions with binary variables that activate discrete control segments. 
While this ensures tighter coordination, the resulting mixed-integer structure can become difficult to scale for large or time-coupled systems \cite{wagle2024optimal,aboshady2023reactive}. 
A complementary approach replaces discontinuous controls with smooth, first-order-continuous relaxations to improve Newton convergence without introducing binary variables \cite{turner2021analytical}. 
Unlike these external or partially integrated formulations, the proposed TSBI model embeds control behavior directly within the converter’s algebraic equations, ensuring that control actions, electrical states, and loss mechanisms are solved implicitly in a single, unified formulation.

\end{added}

\subsection{Proposed Solution and Contributions}
\noindent To address the limitations of existing inverter models, this paper develops an equivalent circuit-based steady-state model of a \Inverter{} (\InverterShort{}) from first principles. 
% —switching and conduction—
The model captures the losses using closed-form semiconductor-based expressions embedded directly into modular circuit equivalents, preserving physical interpretability and generalizability without the need for calibration. 
The model supports bidirectional operation without mode-specific constraints, and the internal losses are formulated using sign-aware expressions that yield the appropriate loss polarity for each direction of power flow.
The model also integrates any inverter control behavior (e.g., maximum power point tracking (MPPT), Volt-Var) without loss of generality by introducing control-dependent controlled current sources. 
These controlled-source expressions are twice differentiable, ensuring compatibility with gradient-based solvers and scalability to large distribution grid optimizations.
The paper addresses significant gaps in inverter modeling by introducing the following novelties:
\begin{itemize}[leftmargin=0.75em]
    \item We develop a physics-based inverter model that captures switching and conduction losses using equivalent circuit components derived from closed-form semiconductor expressions.
    
    \item We incorporate both charging and discharging behavior without the need for binary variable-based switching logic or complementarity-based mode-specific rules.

    \item We model control logic generically into the inverter secondary with controlled sources, as continuous, twice-differentiable functions, enabling direct integration into large-scale nonlinear optimization formulations.
\end{itemize}

\begin{added}{Reviewer 3 Comment 2}
\noindent The remainder of the paper is organized as follows. Section \ref{preliminaries} reviews the equivalent circuit formulation (ECF) for the grid, battery, and PV systems. Section \ref{formulation} derives the \InverterShort{} model from first-principles. Section \ref{sec:Control} models the \InverterShort{} control within the ECF paradigm. Section \ref{Results} provides validation and large-scale results.

\end{added}

\section{PRELIMINARIES} \label{preliminaries}

%To develop the proposed steady-state \InverterShort{} model, we first have to model the systems it interacts with: the distribution grid, photovoltaic (PV) source, and battery storage. These subsystems determine the \InverterShort{}’s input and output conditions and must be modeled consistently. We adopt the equivalent circuit modeling (ECM) framework, which expresses each component using current-voltage (\textit{I--V}) relationships under steady-state conditions. This approach ensures physical accuracy and allows integration into power flow and optimization formulations.
%Inverters interconnect DERs such as rooftop solar and battery systems with the distribution grid.
% \noindent We adopt the equivalent circuit approach to develop the proposed inverter model for large-scale simulation and optimization studies. 
% The remainder of this section therefore reviews equivalent-circuit models for distribution grids, photovoltaics, and batteries.

\noindent We adopt the equivalent circuit approach to develop the proposed inverter model for large-scale simulation and optimization studies. Since the inverter operates between the grid and DC subsystems (battery and PV), this section reviews equivalent circuit models for these connected subsystems.

\begin{added}{Reviewer 2 Comment 6}
\subsection{Equivalent Circuit Model for Distribution Grid Power Flow} \label{sec:Grid_formulation}

\noindent The equivalent circuit model (ECM) framework models three-phase distribution grids through the components' steady-state current-voltage (\textit{I--V}) relationships. 
It formulates the grid as an undirected graph $\mathcal{G}(\mathcal{N}, \mathcal{E})$ with voltages and currents expressed in rectangular coordinates \cite{pandey2018robust, jereminov2016equivalent}. Each phase $\phi, \gamma \in \{a, b, c\}$ is decomposed into real and imaginary parts. Kirchhoff’s Current Law (KCL) is applied at every node $i$ and phase $\phi$, yielding algebraic equations:\end{added} 
\begin{subequations}
\begin{align}
\sum_{j \in \mathcal{N}} \sum_{\gamma \in \Phi} \left( G_{ij}^{\phi\gamma} V_{ij}^{R,\gamma} - B_{ij}^{\phi\gamma} V_{ij}^{I,\gamma} \right) + I_{i}^{R,\phi} &= 0 \label{eq:kcl_real} \\
\sum_{j \in \mathcal{N}} \sum_{\gamma \in \Phi} \left( G_{ij}^{\phi\gamma} V_{ij}^{I,\gamma} + B_{ij}^{\phi\gamma} V_{ij}^{R,\gamma} \right) + I_{i}^{I,\phi} &= 0 \label{eq:kcl_imag}
\end{align}

\noindent where $V_{ij}^{R,\gamma}$ and $V_{ij}^{I,\gamma}$ denote the real and imaginary parts of voltage differences across nodes $i$ and $j$, $I_{i}^{R,\phi}$ and $I_{i}^{I,\phi}$ are the real and imaginary current injections, and $G_{ij}^{\phi\gamma}$, $B_{ij}^{\phi\gamma}$ are the conductance and susceptance terms between phases. Current injections from loads and generators introduce nonlinearities and are modeled for component $k$ as:
% \begin{equation}
% \begin{aligned}
% I_{k,i}^{R,\phi} &= \frac{P_{k,i}^{\phi}\!V_{i}^{R,\phi} \!+\! Q_{k,i}^{\phi}\!V_{i}^{I,\phi}}{(V_{i}^{R,\phi})^2 \!+\! (V_{i}^{I,\phi})^2},
% I_{k,i}^{I,\phi} = \frac{P_{k,i}^{\phi}\!V_{i}^{I,\phi} \!-\! Q_{k,i}^{\phi}\!V_{i}^{R,\phi}}{(V_{i}^{R,\phi})^2 \!+\! (V_{i}^{I,\phi})^2}
% \end{aligned}
% \label{eq:current_injection_compact}
% \end{equation}
{\relsize{-0.5}
\begin{equation}
\begin{aligned}
I_{k,i}^{R,\phi} = \frac{P_{k,i}^{\phi}\!V_{i}^{R,\phi} \!+\! Q_{k,i}^{\phi}\!V_{i}^{I,\phi}}{(V_{i}^{R,\phi})^2 \!+\! (V_{i}^{I,\phi})^2}, 
I_{k,i}^{I,\phi} = \frac{P_{k,i}^{\phi}\!V_{i}^{I,\phi} \!-\! Q_{k,i}^{\phi}\!V_{i}^{R,\phi}}{(V_{i}^{R,\phi})^2 \!+\! (V_{i}^{I,\phi})^2}
\end{aligned}
\label{eq:current_injection_compact}
\end{equation}
}

% \noindent The  net current injections $ I_{i}^{R,\phi} $ and $ I_{i}^{I,\phi} $ at node $i$ and phase $\phi$ is the sum of all load $\mathcal{L}$ and generator $\mathcal{G}$ contributions:

\noindent At each node $i$ and phase $\phi$, the net current injection is computed as the sum of load contributions minus the sum of generator contributions:
% \begin{equation}
% \begin{aligned}
% I_{i}^{R,\phi} &= \sum_{k \in \mathcal{L}_i}\! I_{k,i}^{R,\phi} \!-\! \sum_{m \in \mathcal{G}_i}\! I_{m,i}^{R,\phi},
% I_{i}^{I,\phi} = \sum_{k \in \mathcal{L}_i}\! I_{k,i}^{I,\phi} \!-\! \sum_{m \in \mathcal{G}_i}\! I_{m,i}^{I,\phi}
% \end{aligned}
% \label{eq:current_balance_compact}
% \end{equation}
{\relsize{-0.5}
\begin{equation}
I_{i}^{R,\phi} = \sum_{k \in \mathcal{L}_i}\! I_{k,i}^{R,\phi} \!-\! \sum_{m \in \mathcal{G}_i}\! I_{m,i}^{R,\phi},
I_{i}^{I,\phi} = \sum_{k \in \mathcal{L}_i}\! I_{k,i}^{I,\phi} \!-\! \sum_{m \in \mathcal{G}_i}\! I_{m,i}^{I,\phi}
\label{eq:current_balance_compact}
\end{equation}
}

\end{subequations}

% \noindent Prior work has demonstrated ECM's accuracy and scalability in grid analysis and optimization \cite{ali2024distributed, 10886763}.

\noindent ECM has shown scalable performance in grid studies \cite{ali2024distributed,badmus2024anoca}.

% \noindent
% Current injections come from multiple sources, so it is necessary to model individual DERs. The next section presents equivalent circuit models for the battery and PV.

\begin{figure}[htpb]
    \centering
    \includegraphics[width=1\linewidth]{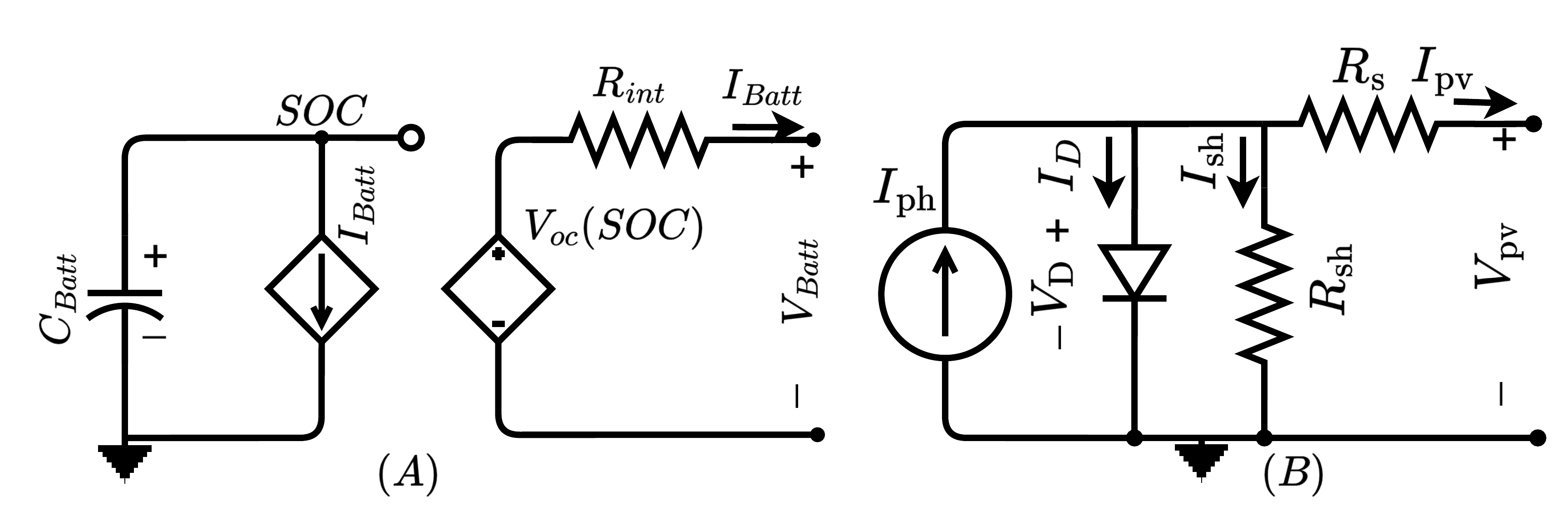}
    \caption{ECF models for: (A) battery (zeroth-order) and (B) PV (SDM)}
    \label{fig:ecf}
    \vspace{0.5em}
\end{figure}

\begin{added}{Reviewer 2 Comment 6}
\subsection{Equivalent Circuit Model for Battery} \label{sec:battery_iv_formulation}
\noindent We use a zeroth-order equivalent circuit model for the battery, reduced from the second-order model in \cite{chen2006accurate}. 
The original model includes a lifetime subcircuit with $R_{\text{SD}}$, $C_{\text{Batt}}$, and $V_{\text{OC}}(SOC)$ in parallel, and a dynamic subcircuit with $R_S$ in series with two RC branches. 
Assuming $R_{\text{SD}} \to \infty$, we approximate the SOC dynamics with trapezoidal integration:
\end{added}
\begin{subequations}
\begin{equation}
SOC(t{+}1) = SOC(t) - \frac{\Delta t}{2 C_{\text{Batt}}} \left( I_{\text{Batt}}(t) + I_{\text{Batt}}(t{+}1) \right)
\label{eq:SOC_update}
\end{equation}

% We assume $\Delta t \gg \tau_s, \tau_l$, where $\tau_s = R_{\text{Ts}} C_{\text{Ts}}$ and $\tau_l = R_{\text{Tl}} C_{\text{Tl}}$. This allows us to neglect both short and long-term capacitor dynamics and use a steady-state model. The terminal voltage becomes

\noindent We also assume $\Delta t \gg \tau_s, \tau_l$, where $\tau_s = R_{\text{Ts}} C_{\text{Ts}}$ and $\tau_l = R_{\text{Tl}} C_{\text{Tl}}$ are time constants representing the short and long term transient response of the battery. This allows us to neglect both responses in the steady-state model. The terminal voltage becomes:
\begin{equation}
V_{\text{Batt}}(t) = V_{\text{OC}}(SOC(t)) - I_{\text{Batt}}(t) R_{\text{int}}
\label{eq:battery_voltage}
\end{equation}
\end{subequations}

\noindent where $R_{\text{int}} = R_S + R_{Ts} + R_{Tl}$. Fig. \ref{fig:ecf}(A) shows the final equivalent circuit used in the studies.
% \begin{figure}[htpb]
%     \centering
%     \includesvg[width=1\linewidth]{images/Batt_PV_ECF.drawio.svg}
%     \caption{ECF models for: (A) battery (zeroth-order) and (B) PV (SDM)}
%     \label{fig:ecf}
%     \vspace{0.5em}
% \end{figure}

\subsection{Equivalent Circuit Model for PV} \label{sec:pv_iv_formulation}

\noindent We model the electrical behavior of photovoltaic (PV) systems using the single-diode model (SDM) \cite{phang1984accurate}, which captures the nonlinear characteristics of the p-n junction. The equivalent circuit includes a photocurrent source, diode, series resistance $R_s$, and shunt resistance $R_{\text{sh}}$ (see Fig. \ref{fig:ecf}(B)). KCL and the Shockley equation yield the output current:
\begin{equation}
I_{\text{pv}} = I_{\text{ph}} - I_0 \left( e^{\frac{V_{\text{pv}} + I_{\text{pv}} R_s}{n_d N_s V_t}} - 1 \right) - \frac{V_{\text{pv}} + I_{\text{pv}} R_s}{R_{\text{sh}}}
\label{eq:pv_iv_final}
\end{equation}

\noindent where $I_{\text{pv}}$, $V_{\text{pv}}$, and $I_{\text{ph}}$ denote the output current, terminal voltage, and photocurrent; $I_0$ is the reverse saturation current; $R_s$ and $R_{\text{sh}}$ are series and shunt resistances; $n_d$ is the diode factor; and $V_t$ is the thermal voltage per cell out of $N_s$ series cells.

%With the defined grid, battery, and PV models, we now formulate the \InverterShort{}{} model in the next section.

\section{\InverterShort{} Model Formulation} \label{formulation}

\noindent We model the \InverterShort{} as a four-block system comprising a four-switch non-inverting buck-boost converter, referred to as the first-stage converter (\FSC{}); a DC-link filter; a single-phase H-bridge converter, referred to as the second-stage converter (\SSC{}); and an AC filter. The \FSC{} uses duty-cycle modulation to adjust variable DER voltages to a regulated DC-link level for both forward and reverse power flow. The DC-link filter attenuates switching ripple and stabilizes the intermediate voltage, typically using capacitors and inductors whose configuration depends on converter design and ripple requirements. 
We do not explicitly model DC-link in this work, as in DC steady-state, inductors are treated as short circuits, capacitors as open circuits, and we ignore parasitic resistances due to their minimal impact. 
The \SSC{}, driven by sinusoidal PWM, converts power between the DC link and AC terminal in all four quadrants, with the modulation index $m$ governing voltage transformation. 
An AC filter forms the final block and here, it is modeled as an LCL filter operating in steady state at the nominal grid frequency and it attenuates higher switching harmonics.
All blocks are represented using modular equivalent circuits that include relevant losses.

\subsection{First-Stage Converter (\FSC{})} \label{subsec:first_stage_converter_model}

\noindent The \FSC{} consists of four active switches and one inductor configured for non-inverting, bidirectional buck-boost operation, as demonstrated in practical designs such as \cite{TexasInstruments2018}. 
All switches are modulated continuously to avoid discrete mode transitions and support a wide input range \cite{zhang2020advanced}. 
We analyze the converter under forward power flow; reverse flow only differs in current direction due to the symmetric switching pattern.

\begin{figure}[htpb]
    \centering
    \includegraphics[width=1\linewidth]{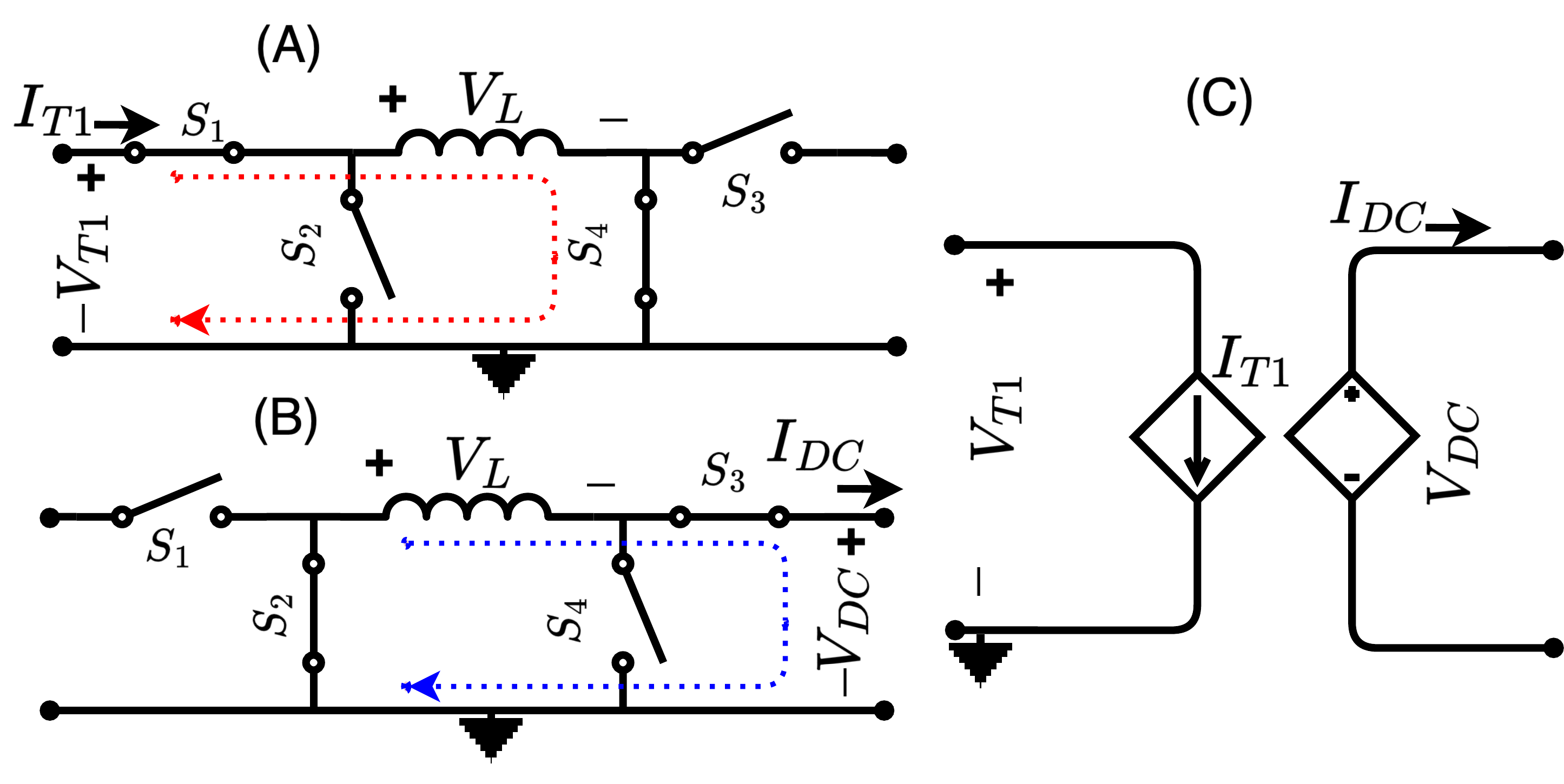}
    \caption{Idealized \FSC{} operation under forward power flow. (A) \FESM{}: inductor charges from $V_{\text{T1}}$ via $S_1$, $S_4$. (B) \FETM{}: energy transferred to $V_{\text{DC}}$ via $S_2$, $S_3$. (C) Equivalent transformer model relating $V_{\text{T1}}$ to $V_{\text{DC}}$.}
    \label{fig:bbc_equivalent}
\end{figure}

Each power flow direction consists of two switching intervals per cycle. 
In the forward case, the converter operates in \ForwardEnergyStorageMode{} (\FESM{}), where a diagonal pair of switches ($S_1$ and $S_4$ in Fig. \ref{fig:bbc_equivalent}A) conduct for $D T_s$ seconds, allowing the inductor to charge at voltage $V_{\text{T1}}$. 
In \ForwardEnergyTransferMode{} (\FETM{}), the other pair ($S_2$ and $S_3$ in Fig. \ref{fig:bbc_equivalent}B) conduct for $(1 - D) T_s$ seconds, discharging the inductor into the DC link at voltage $V_{\text{DC}}$. 
% During reverse mode,\ReverseEnergyStorageMode{} (\RESM{}) and \ReverseEnergyTransferMode{} (\RETM{}), similar switching occurs with reverse current flow.
During reverse mode, \ReverseEnergyStorageMode{} (\RESM{}) and \ReverseEnergyTransferMode{} (\RETM{}) execute the same sequence but with negative current.

% \begin{figure}[htpb]
%     \centering
%     \includesvg[width=1\linewidth]{images/BBC_equivalent_nl.drawio.svg}
%     \caption{Idealized \FSC{} operation under forward power flow. (A) \FESM{}: inductor charges from $V_{\text{T1}}$ via $S_1$, $S_4$. (B) \FETM{}: energy transferred to $V_{\text{DC}}$ via $S_2$, $S_3$. (C) Equivalent transformer model relating $V_{\text{T1}}$ to $V_{\text{DC}}$.}
%     \label{fig:bbc_equivalent}
% \end{figure}

 Assuming continuous conduction mode (CCM), the average voltage across the inductor over a switching cycle is zero, ensuring stable inductor current and preventing drift, and therefore yielding the voltage transformation expression:
\begin{subequations}
\begin{equation}
\frac{1}{T_s} \left( D T_s V_{\text{T1}} - (1 - D) T_s (V_{\text{DC}})\right) = 0  \Rightarrow  
V_{\text{DC}} = \frac{D}{1 - D} \, V_{\text{T1}}
\label{eq:dc_voltage_relation}
\end{equation}

\noindent where duty cycle $D$ determines the operating region: buck ($D < 0.5$), boost ($D > 0.5$), or unity gain ($D = 0.5$). 
%For generality, we define the turns ratio $T = \frac{D}{1 - D}$, yielding $V_{\text{DC}} = T V_{\text{T1}}$, analogous to an ideal transformer. 
To derive current relationships, we use power balance across the converter:
\begin{equation}
V_{\text{T1}} I_{\text{T1}} = V_{\text{DC}} I_{\text{DC}}
\label{eq:power_balance}
\end{equation}
\end{subequations}

 These expressions define a lossless voltage and power transformation, allowing the \FSC{} to be modeled like an ideal transformer. This relationship holds for both forward and reverse modes. 
 The equivalent circuit is shown in Fig. \ref{fig:bbc_equivalent}C. 
 Next, we extend the ideal \FSC{} model by introducing \textit{conduction} and \textit{switching} losses.
 %and derive closed-form expressions to build a non-ideal equivalent circuit for steady-state analysis.

\subsubsection{\FSC{} Switching Losses} \label{subsec:switching_losses}

Switching losses in the \FSC{} result from the overlap between time-varying voltage and current during transistor transitions in each PWM cycle. 
% Each switch undergoes turn-on and turn-off events on both the terminal \TOne{} and DC-link sides. 
% We apply a linear model \cite{mohan2003power}, assuming that current increases or decreases linearly over the switching interval and voltage remains constant.
Per the linear model in \cite{mohan2003power}, we assume that current varies linearly during the switching interval while voltage remains constant.

% The duration for on and off transitions are given by $t_{\text{on}} = t^{\text{on}}_{d} + t_r$ and $t_{\text{off}} = t^{\text{off}}_{d} + t_f$, respectively, where $t^{\text{on}}_{d}$, $t^{\text{off}}_{d}$ are delay times (0--10\%) and $t_r$, $t_f$ are rise/fall times (10--90\%). The remaining is negligible due to near-zero $\frac{dI}{dt}$ and it's always omitted in datasheets as standardized in \cite{jedec1989}. 

\noindent We define the on and off transition durations as:
\begin{subequations}
\begin{equation}
t_{\text{on}}= t_{\text{delay}}^{\text{on}} + t_{\text{rise}}, \quad 
t_{\text{off}} = t_{\text{delay}}^{\text{off}} + t_{\text{fall}}
\end{equation}

\noindent where $t_{\text{delay}}^{\text{on}}$ and $t_{\text{delay}}^{\text{off}}$ denote the initial segments of the current transition during turn-on and turn-off, respectively, and $t_{\text{rise}}$, $t_{\text{fall}}$ correspond to the remaining transition intervals during which the current ramps up or down.
% \noindent where $t_{\text{delay}}^{\text{on}}$ and $t_{\text{delay}}^{\text{off}}$ denote the 0--10\% current intervals, and $t_{\text{rise}}$, $t_{\text{fall}}$ correspond to the 10--90\% rise and fall times. 
% The remaining is negligible due to near-zero $\frac{dI}{dt}$ and it's always omitted in datasheets as standardized in \cite{jedec1989}. 
This yields a triangular power profile, with energy per event given by: 
\begin{equation}
E_{\text{SW}} = \frac{1}{2} V_{\text{SW}} I_{\text{SW}} t_{\text{x}} 
\label{eq:Esw_formula}
\end{equation}

\noindent where $t_x \in \{t_{\text{on}}, t_{\text{off}}\}$. Table \ref{tab:FSC_switching_events_loss_compact} summarizes the switching events and energy losses. 
Each switch undergoes one ON and one OFF transition, resulting in a total of two transitions per switching cycle.

\begin{table}[htpb]
% \caption{Switching Events, Voltages, and Energy Losses per Cycle}
\caption{Switching events and per-cycle energy in the FSC. Each transition dissipates $\tfrac{1}{2}V_{\mathrm{SW}}I_{\mathrm{SW}}t_x$, and two devices switch per sub-interval.}
\label{tab:FSC_switching_events_loss_compact}
\centering
\renewcommand{\arraystretch}{1.1}
\setlength{\tabcolsep}{4pt}
\begin{tabular}{|c|c|l|}
\hline
\textbf{Mode} & \textbf{Voltage} & \textbf{Switching Events and Energy Losses} \\
\hline
\textbf{FESM} & $V_{\text{T1}}$ &
\begin{tabular}[c]{@{}l@{}}
$S_1$, $S_4$ ON:  $0 \to I_{\text{T1}}$ \vrule{}
$S_2$, $S_3$ OFF:  $I_{\text{T1}} \to 0$ \\
\hline
Energy Loss: $2 \cdot \frac{1}{2} V_{\text{T1}} I_{\text{T1}} t_{\text{on}}$ + $2  \cdot \frac{1}{2} V_{\text{T1}} I_{\text{T1}} t_{\text{off}}$
\end{tabular} \\
\hline
\textbf{FETM} & $V_{\text{DC}}$ &
\begin{tabular}[c]{@{}l@{}}
$S_2$, $S_3$ ON:  $0 \to I_{\text{DC}}$ \vrule{}
$S_1$, $S_4$ OFF:  $I_{\text{DC}} \to 0$ \\
\hline
Energy Loss: $2 \cdot \frac{1}{2} V_{\text{DC}} I_{\text{DC}} t_{\text{on}}$ + $2  \cdot \frac{1}{2} V_{\text{DC}} I_{\text{DC}} t_{\text{off}}$
\end{tabular} \\
\hline
\end{tabular}
\end{table}

% Switching transitions repeat in both FESM and FETM intervals, 
% making the total energy dissipated per switching cycle from Table \ref{tab:FSC_switching_events_loss_compact} to be:
% \noindent The expressions in the last column of Table \ref{tab:FSC_switching_events_loss_compact} are obtained directly from the linear overlap model in \eqref{eq:Esw_formula}, where each switching transition dissipates an energy of $\tfrac{1}{2}V_{\mathrm{SW}} I_{\mathrm{SW}} t_x$. Two devices undergo transitions within each sub-interval—one turning ON and one turning OFF—resulting in the factor $2 \cdot \tfrac{1}{2}$ in the per-cycle energy expressions. Substituting $(V_{\mathrm{SW}}, I_{\mathrm{SW}})=(V_{\mathrm{T1}}, I_{\mathrm{T1}})$ for the FESM interval and $(V_{\mathrm{SW}}, I_{\mathrm{SW}})=(V_{\mathrm{DC}}, I_{\mathrm{DC}})$ for the FETM interval, and summing both contributions over a complete switching cycle yields the total switching energy in \eqref{eq:Esw_total

\begin{added}{Reviewer 3 Comment 3}
\noindent The energy loss expressions in Table \ref{tab:FSC_switching_events_loss_compact} follow directly from the linear overlap model in \eqref{eq:Esw_formula}, where each switching transition dissipates an energy of $\tfrac{1}{2} V_{\text{SW}} I_{\text{SW}} t_x$. 
Two devices switch within each sub-interval, one turning ON and one turning OFF, giving the factor $2 \cdot \tfrac{1}{2}$ in the per-cycle energy terms. 
Substituting $(V_{\text{SW}}, I_{\text{SW}}) \equiv (V_{T1}, I_{T1})$ for the FESM interval and $(V_{\text{SW}}, I_{\text{SW}}) \equiv (V_{DC}, I_{DC})$ for the FETM interval, and summing both contributions over a full switching cycle yields the total switching energy in \eqref{eq:Esw_total}.

\end{added}
\begin{equation}
E_{\text{SW,total}} = (t_{\text{on}} + t_{\text{off}}) \left( 2 \cdot \tfrac{1}{2} V_{\text{T1}} I_{\text{T1}} + 2 \cdot \tfrac{1}{2} V_{\text{DC}} I_{\text{DC}} \right)
\label{eq:Esw_total}
\end{equation}

\noindent Multiplying \eqref{eq:Esw_total} by the \FSC{} switching frequency $f_{\text{sw}}^{1}$ yields the average switching-loss power, and partitioning it into terminal-specific contributions gives:
\begin{equation}
P^{T1}_{\text{SW}} \!=\! f^1_{\text{sw}} (t_{\text{on}}\!+\!t_{\text{off}}) V_{\text{T1}} I_{\text{T1}},
P^{DC}_{\text{SW}} \!=\! f^1_{\text{sw}} (t_{\text{on}}\!+\!t_{\text{off}}) V_{\text{DC}} I_{\text{DC}}
\label{eq:switching_losses}
\end{equation}

We then represent these switching losses with controlled current sources whose expressions follow \eqref{eq:switching_current_t1_sgn} and \eqref{eq:switching_current_dc_sgn}.
% We then use controlled current sources to model the switching losses in the equivalent circuit with expressions defined in \eqref{eq:switching_current_t1_sgn} and \eqref{eq:switching_current_dc_sgn}.
We obtain these expressions by dividing the power terms by the voltage across the switch. The $\operatorname{sgn}(\cdot)$ operator ensures strictly positive loss power for either current direction.
%These loss currents are injected in series with each terminal current in the direction of power flow, increasing the source-side draw and reducing the net terminal output.
\begin{align}
I^{T1}_{\text{SW}} \!&=\! f^1_{\text{sw}}(t_{\text{on}}\!+\!t_{\text{off}}) \operatorname{sgn}(I_{\text{T1}}) I_{\text{T1}} \!=\! f^1_{\text{sw}}(t_{\text{on}}\!+\!t_{\text{off}}) |I_{\text{T1}}| \label{eq:switching_current_t1_sgn} \\
I^{DC}_{\text{SW}} \!&=\! f^1_{\text{sw}}(t_{\text{on}}\!+\!t_{\text{off}}) \operatorname{sgn}(I_{\text{DC}}) I_{\text{DC}} \!=\! f^1_{\text{sw}}(t_{\text{on}}\!+\!t_{\text{off}}) |I_{\text{DC}}| \label{eq:switching_current_dc_sgn}
\end{align}
\end{subequations}

%During forward flow ($I_{\text{DC}} > 0$), $I^{T1}_{\text{SW}}$ reduces the DC-link output while $I^{DC}_{\text{SW}}$ increases the terminal \TOne{} current drawn. The directions reverse for $I_{\text{DC}} < 0$, maintaining energy balance under all conditions.

\subsubsection{\FSC{} Conduction Losses} \label{subsec:conduction_losses}
Conduction losses in the \FSC{} arise from voltage drops across the conducting switches and the inductor’s parasitic resistance. 
We model voltage across a conducting switch with a linear relation $v = V_{T0} + R i$, where $V_{T0}$ and $R$ denote the threshold voltage and on-state resistance, respectively \cite{mohan2003power}. 
%Under steady-state conditions, the average voltage across the inductor is zero. 
To derive expressions for conduction losses, we apply KVL to each conduction path (\FESM{} and \FETM{}) in Fig. \ref{fig:forward_power_transfer} and aggregate following the superposition theorem. 
Applying over the switching cycle yields a composite voltage relationship:
%based on KVL applied to both \FESM{} and \FETM{} paths.
\begin{subequations}
\begin{equation}
\begin{aligned}
D V_{\text{T1}} - (1 - D) V_{\text{DC}} &=
D (2 V_{T0} + I_{\text{T1}} (2 R_T + R_L)) \\
&\hspace{-2.7em}+ (1 - D) (2 V_{T0} + I_{\text{DC}} (2 R_T + R_L))
\end{aligned}
\end{equation}
% \begin{figure}[htbp]
%     \centering
%     \includesvg[width=0.85\linewidth]{images/BBC_equivalent_nl_w_res.drawio.svg}
%     \caption{ (A) \FESM{} with current $I_{\text{T1}}$, duty $D$; (B) \FETM{} with current $I_{\text{DC}}$, duty $1{-}D$. Losses include $V_{T0}$, $R_T$, and $R_L$, and the total voltage drop is obtained by superposition of the two subcircuits.}
%     \label{fig:forward_power_transfer}
% \end{figure}
\begin{figure}[htbp]
    \centering
    \includegraphics[width=0.85\linewidth]{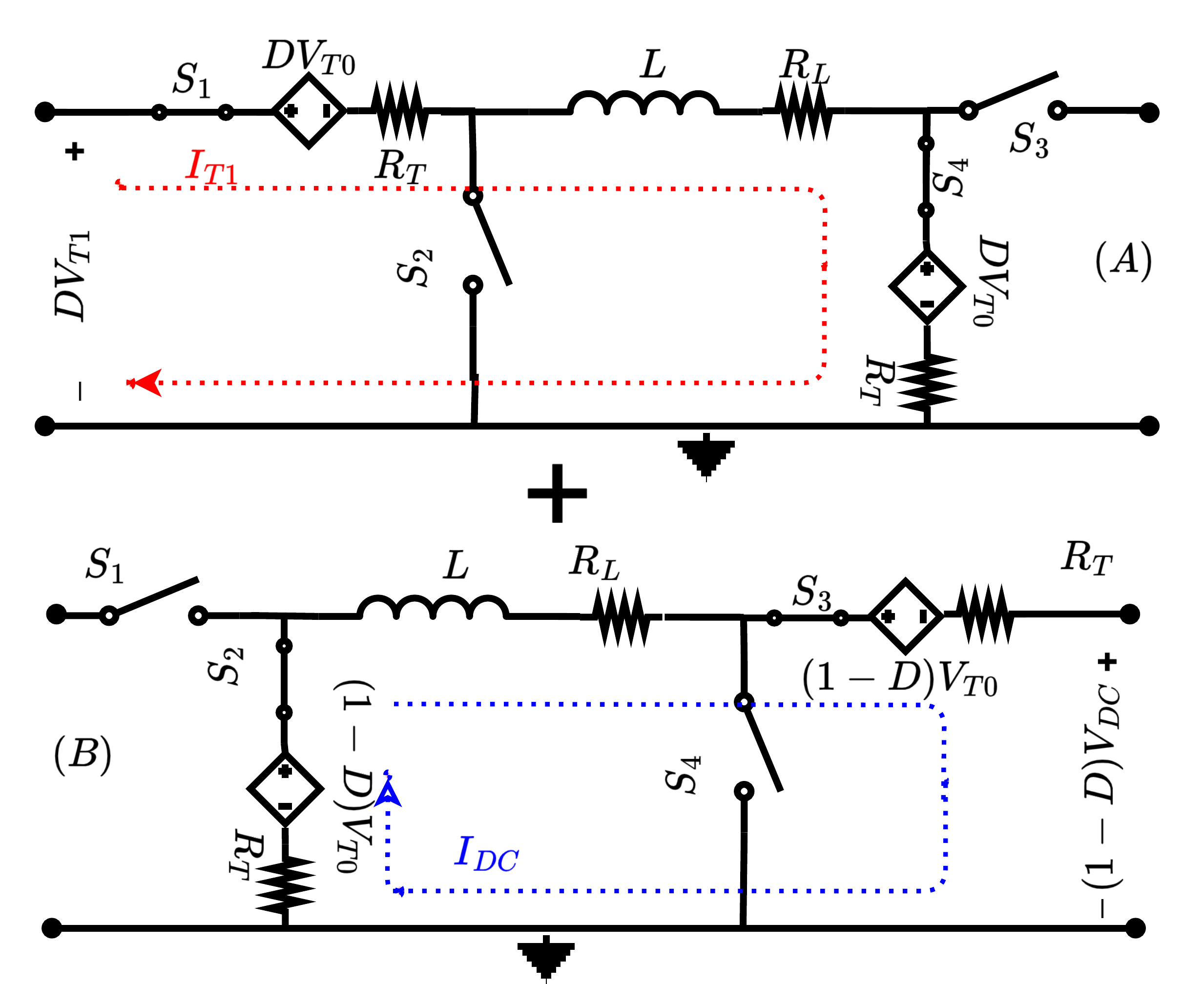}
    \caption{(A) \FESM{} with current $I_{\text{T1}}$, duty $D$; (B) \FETM{} with current $I_{\text{DC}}$, duty $1{-}D$. Losses include $V_{T0}$, $R_T$, and $R_L$, and the total voltage drop is obtained by superposition of the two subcircuits.}
    \label{fig:forward_power_transfer}
\end{figure}

The left-hand side matches the ideal voltage relation \eqref{eq:dc_voltage_relation}, while the right-hand side captures conduction losses as voltage drops.
The ideal term remains unchanged under reverse flow, whereas the loss terms change sign with current polarity.
To generalize the forward and reverse model, we apply the sign function to the threshold voltage terms and express the total conduction drop as:
\begin{equation}
\begin{aligned}
V_C &= D \left( 2\, \operatorname{sgn}(I_{\text{T1}})\, V_{T0} + I_{\text{T1}} (2R_T + R_L) \right) \\
&\quad + (1 - D) \left( 2\, \operatorname{sgn}(I_{\text{DC}})\, V_{T0} + I_{\text{DC}} (2R_T + R_L) \right)
\end{aligned}
\end{equation}

\noindent We implement these drops as controlled voltage sources with:
\begin{align}
V^{T1}_{\text{C}} &= D \left( 2 \operatorname{sgn}(I_{\text{T1}})\, V_{T0} + I_{\text{T1}} (2R_T + R_L) \right) \label{eq:VC_T1} \\
V^{DC}_{\text{C}} &= (1 - D) \left( 2 \operatorname{sgn}(I_{\text{DC}})\, V_{T0} + I_{\text{DC}} (2R_T + R_L) \right) \label{eq:VC_DC}
\end{align}
\end{subequations}

The $\operatorname{sgn}(\cdot)$ operator ensures that the threshold voltage contributes a positive power loss regardless of flow direction.

% \begin{figure}[htpb]
%     \centering
%     \includesvg[width=0.75\linewidth]{images/BBC_equivalent.drawio.svg}
%     \caption{Equivalent circuit model of non-ideal \FSC{}, with switching and conduction losses using controlled sources.}
%     \label{fig:first_stage_converter_equiv}
% \end{figure}
\begin{figure}[htpb]
    \centering
    \includegraphics[width=0.75\linewidth]{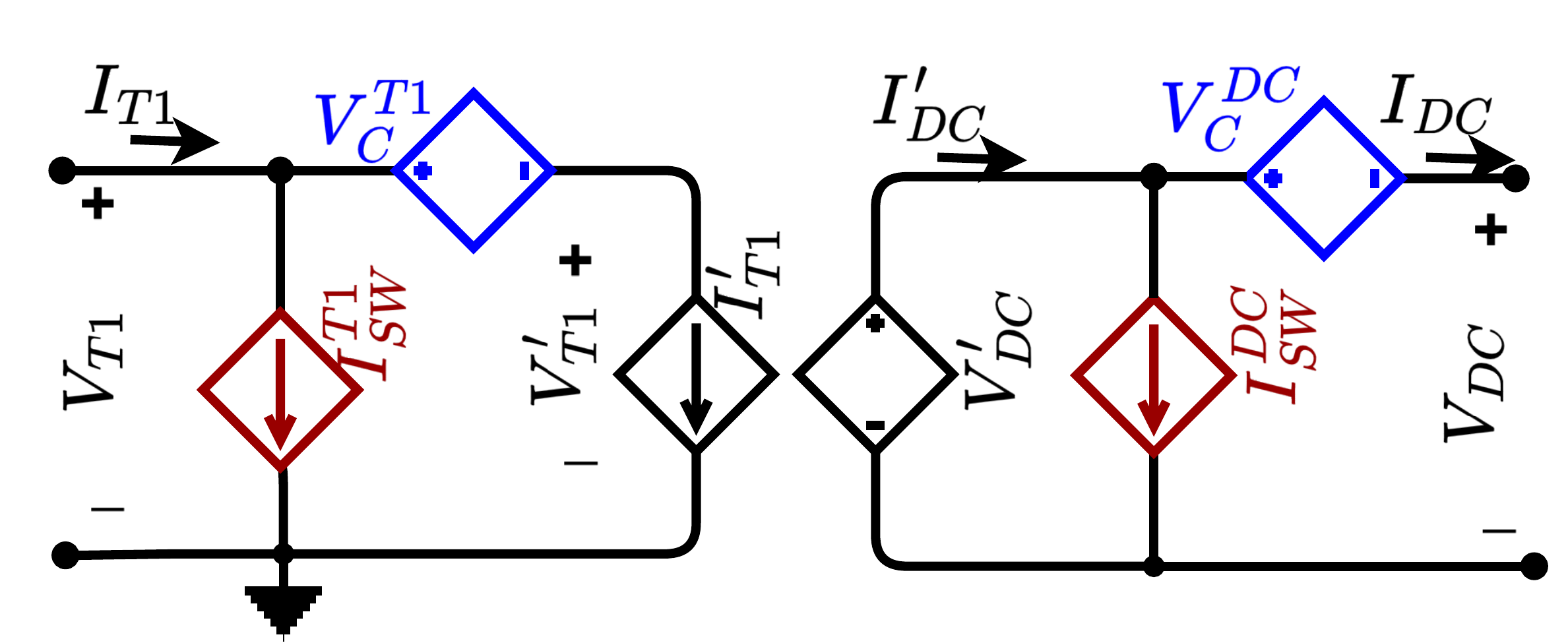}
    \caption{Equivalent circuit model of non-ideal \FSC{}, with switching and conduction losses using controlled sources.}
    \label{fig:first_stage_converter_equiv}
\end{figure}

\noindent Fig. \ref{fig:first_stage_converter_equiv} shows the non-ideal \FSC{} ECM, where controlled sources add switching and conduction losses.

\begin{remark}
    \noindent The sign function $\operatorname{sgn}(x) = x/|x|$ governs direction-dependent losses. 
    We approximate $|x|$ with $\sqrt{x^2 + \epsilon}$, yielding:
    \begin{subequations}
    \begin{equation}
    \operatorname{sgn}(x) \approx \frac{x}{\sqrt{x^2 + \epsilon}}, \quad \epsilon > 0
    \label{eq:sgn_abs}
    \end{equation}
    
    \noindent where $\epsilon > 0$ is a small constant added to avoid division by zero and to ensure smoothness near the origin. This form is continuously differentiable over $\mathbb{R}$, with derivative:
    \begin{equation}
    \frac{d}{dx} \operatorname{sgn}(x) \approx \frac{\epsilon}{(x^2 + \epsilon)^{3/2}}
    \end{equation}
    \end{subequations}
    
    \noindent which is bounded and Lipschitz continuous \cite{amhraoui2022smoothing}. The smooth form facilitates solver stability and enables gradient-based optimization in this paper.
\end{remark}

% \subsubsection{Integration of Losses into the Non-Ideal \FSC{} Circuit Model}

% Using controlled sources, we integrate switching and conduction losses in the \FSC{} circuit, as shown in Fig. \ref{fig:first_stage_converter_equiv}. 
% Switching losses, derived in \eqref{eq:switching_current_t1_sgn}–\eqref{eq:switching_current_dc_sgn}, are implemented as controlled current sources that inject $I^{T1}_{\text{SW}}$ and $I^{DC}_{\text{SW}}$ at the terminals:
% \begin{subequations}
% \begin{align}
%  I_{\text{T1}} - I_{\text{T1}}' - I^{T1}_{\text{SW}} = 0, \quad
%  I_{\text{DC}}' - I_{\text{DC}} - I^{DC}_{\text{SW}} = 0
% \end{align}

% \noindent where $I_{\text{T1}}'$ and $I_{\text{DC}}'$ are the ideal terminal currents that now replace $I_{\text{T1}}$ and $I_{\text{DC}}$ in the ideal power balance equation \eqref{eq:power_balance}. 

% Conduction losses, based on \eqref{eq:VC_T1}–\eqref{eq:VC_DC}, are modeled as controlled voltage sources in series with each terminal. 
% Applying KVL, the modified terminal voltages become:
% \begin{align}
%  -V_{\text{T1}}' + V_{\text{T1}} - V^{T1}_{\text{C}} = 0, \quad
%  V_{\text{DC}}' - V_{\text{DC}} - V^{DC}_{\text{C}} = 0
% \end{align}
% \end{subequations}

% \noindent where $V_{\text{T1}}'$ and $V_{\text{DC}}'$ are the ideal terminal voltages that now replace $V_{\text{T1}}$ and $V_{\text{DC}}$ in the ideal voltage transformation relation \eqref{eq:dc_voltage_relation} and \eqref{eq:power_balance}.

\subsection{Second-Stage Converter Model (\SSC{})}
\label{subsec:second_stage_converter_model}

% \noindent We model the \SSC{} as a bidirectional single-phase H-bridge inverter regulated by unipolar sinusoidal PWM. Under ideal operation, the output voltage follows a sinusoidal waveform with an RMS value.
% \begin{subequations}
% \begin{equation}
% \vec{V}_{\text{AC}} = \frac{M V_{\text{DC}}}{\sqrt{2}} \angle \theta_V
% \label{eq:AC_RMS}
% \end{equation}

% \noindent where $M$ is the modulation index and $\theta_V$ is the phase angle. Expressed in rectangular form, the real and imaginary voltage components are:
% \begin{equation}
% V_{\text{AC}}^R = \frac{M^R V_{\text{DC}}}{\sqrt{2}}, \quad V_{\text{AC}}^I = \frac{M^I V_{\text{DC}}}{\sqrt{2}}
% \label{eq:SSC_voltage_components}
% \end{equation}

% \noindent with $M^R = M \cos\theta_V$ and $M^I = M \sin\theta_V$. Since we are assuming a lossless operation, the inverter behaves as a modulation-controlled transformer with turns ratio $T = \frac{M}{\sqrt{2}}$, the power transferred between the DC and AC terminals satisfies:
% \begin{equation}
% V_{\text{DC}} I_{\text{DC}} = \Re( \vec{V}_{\text{AC}} \vec{I}_{\text{AC}}^* )
% = |\vec{V}_{\text{AC}}| |\vec{I}_{\text{AC}}| \cos(\phi)
% = V_{\text{AC}}^{R} I_{\text{AC}}^{R} + V_{\text{AC}}^{I} I_{\text{AC}}^{I}
% \label{eq:ssc_power_balance}
% \end{equation}
% \end{subequations}

\noindent We model the \SSC{} as a bidirectional single-phase H-bridge inverter driven by unipolar sinusoidal PWM. Under ideal operation, the fundamental output voltage has an RMS magnitude:
\begin{subequations}
\begin{equation}
|V_{\text{AC}}| = \frac{M V_{\text{DC}}}{\sqrt{2}}
\label{eq:AC_RMS_mag}
\end{equation}

% \noindent where $M \in [0, 1]$ is the scalar modulation index.      In phasor form, the relationship between AC and DC voltages is given by the following expression:

\noindent where $M\in[0, 1]$ is the scalar modulation index. In phasor form, the AC-to-DC voltage relation is:
\begin{equation}
\mathbf{V}_{\text{AC}} = \frac{V_{\text{DC}}}{\sqrt{2}}\, \mathbf{M} 
= \frac{M V_{\text{DC}}}{\sqrt{2}} \angle \theta_V
\label{eq:AC_RMS}
\end{equation}

\noindent where $\mathbf{M} = M \angle \theta_V$ is the complex modulation phasor. Expressing the AC voltage in rectangular form, the real and imaginary voltage components are:
\begin{equation}
V_{\text{AC}}^R = \frac{M^R V_{\text{DC}}}{\sqrt{2}}, \quad 
V_{\text{AC}}^I = \frac{M^I V_{\text{DC}}}{\sqrt{2}}
\label{eq:SSC_voltage_components}
\end{equation}

\noindent with $M^R = M \cos\theta_V$ and $M^I = M \sin\theta_V$. Under the lossless assumption, the inverter behaves as a modulation-controlled transformer with effective turns ratio $T = \frac{M}{\sqrt{2}}$ and phase shift of $\theta_V$. 
The power balance between the DC and AC sides enforces:
\begin{equation}
V_{\text{DC}} I_{\text{DC}} 
%=\Re( \mathbf{V}_{\text{AC}}\, \mathbf{I}_{\text{AC}}^* )
= |\mathbf{V}_{\text{AC}}|\, |\mathbf{I}_{\text{AC}}| \cos(\phi)
= V_{\text{AC}}^{R} I_{\text{AC}}^{R} + V_{\text{AC}}^{I} I_{\text{AC}}^{I}
\label{eq:ssc_power_balance}
\end{equation}
\end{subequations}

\begin{figure}[htpb]
    \centering
    \includegraphics[width=0.7\linewidth]{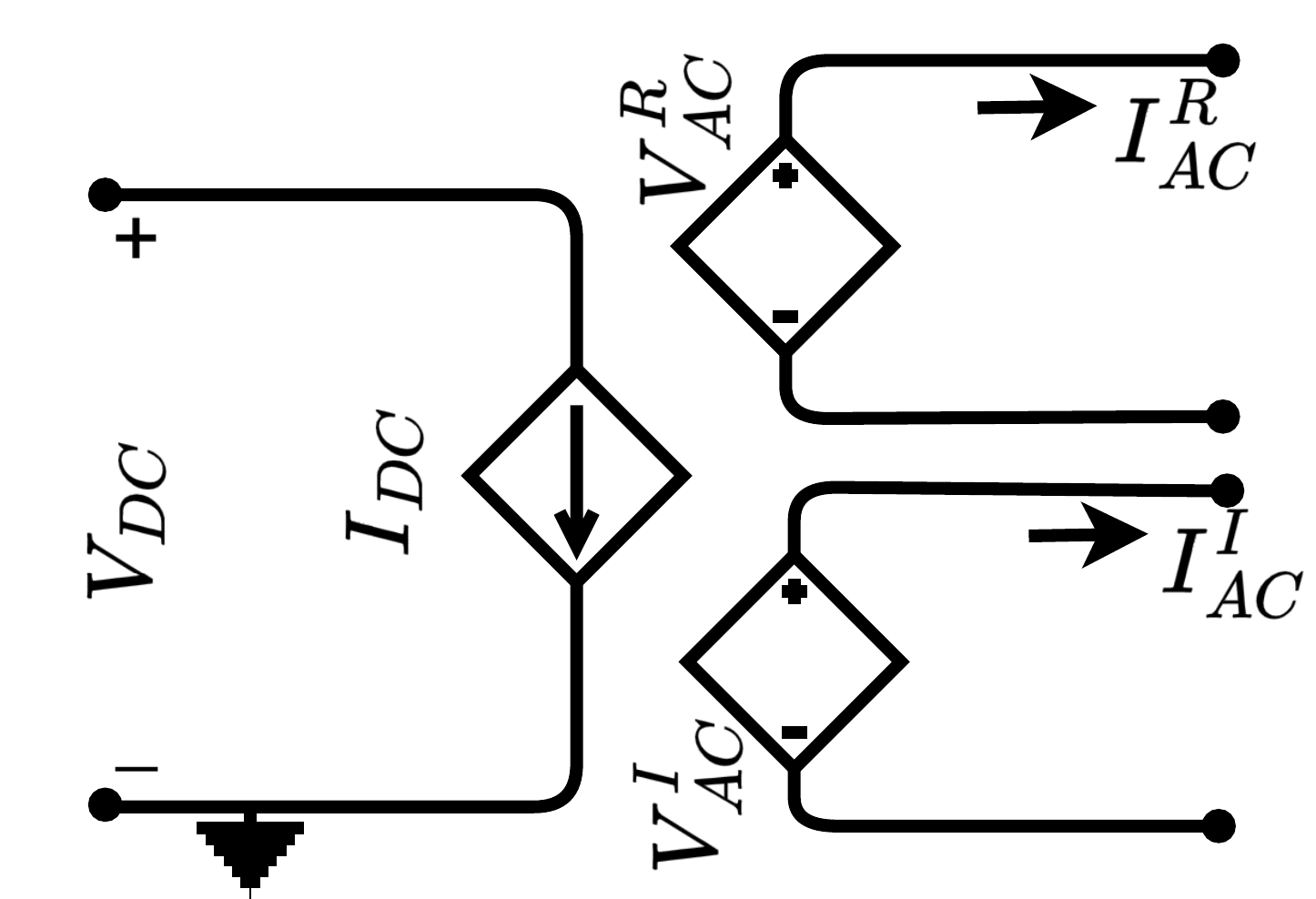}
    \caption{Equivalent circuit representation of the ideal \SSC{}.}
    \label{fig:SSC_equivalent}
\end{figure}

%\noindent This relation enforces real power balance via the complex power formulation. As illustrated in 
\noindent In Fig. \ref{fig:SSC_equivalent}, the ideal \SSC{} maps the DC terminal to real and imaginary secondary terminals. 
We include losses next.
%The structure functions as a phasor-domain transformer controlled by the modulation index under bidirectional operation.

% \begin{figure}[htpb]
%     \centering
%     \includesvg[width=0.7\linewidth]{images/SSC_equivalent_nl3.drawio.svg}
%     \caption{Equivalent circuit representation of the ideal \SSC{}.}
%     \label{fig:SSC_equivalent}
% \end{figure}

\subsubsection{\SSC{} Switching Losses}

\noindent Switching losses in the \SSC{} result from transistor transitions and diode reverse recovery. Just like the \FSC{}, we use the linear energy loss model assuming the voltage across the switch remains constant and current varies linearly during the transition interval \cite{mohan2003power}.

\noindent \textit{a) Transistor Switching Losses:} Under unipolar SPWM, each transistor experiences a voltage swing of $\pm V_{\text{DC}}/2$ and conducts current for one half of the AC cycle. 
The instantaneous AC current is:
\begin{subequations}
\begin{equation}
i(\theta) = \sqrt{2} \, |\mathbf{I}_{\text{AC}}| \sin(\theta + \theta_i)
\label{AC_current_inst}
\end{equation}

\noindent where $\theta = \omega t$ is referenced to the current phasor. The average AC current during any half-cycle of conduction is defined as:
\begin{equation}
\bar{I}_{\text{AC}} = \frac{2\sqrt{2}}{\pi} |\mathbf{I}_{\text{AC}}|
\label{eq:IAC_avg}
\end{equation}

\noindent Using the linear energy loss model, as in \FSC{}, the energy loss per switching event per transistor is:
\begin{equation}
E_{\text{sw}} = \frac{1}{2}  \frac{V_{\text{DC}}}{2}  \frac{2\sqrt{2}}{\pi} |\mathbf{I}_{\text{AC}}| \, t_{\text{sw}}
\end{equation}

% \noindent where $\theta = \omega t$ is referenced to the current phasor and the average current over any half-cycle of conduction is $\frac{2\sqrt{2}}{\pi} |\mathbf{I}_{\text{AC}}|$. Using the linear energy loss model, as in \FSC{}, the energy loss per switching event is:

% \noindent with only two transistors switching (on and off) in each half cycle, each switching at $f_{\text{sw}}$, the resulting switching power loss is:

% \begin{equation}
% P_{\text{sw}} = 8 f_{\text{sw}} E_{\text{sw}} = \frac{2\sqrt{2}}{\pi} f_{\text{sw}} V_{\text{DC}} |\mathbf{I}_{\text{AC}}| (t_{\text{on}} + t_{\text{off}})
% \label{eq:P_sw_bridge}
% \end{equation}

% \noindent where $t_{\text{on}} = t^{\text{on}}_d + t_r$ and $t_{\text{off}} = t^{\text{off}}_d + t_f$ are the effective turn-on and turn-off times, respectively.

\noindent with only two transistors active in each half cycle, each undergoes one turn-on and one turn-off transition per switching event. Since there are two transistors per half-cycle and two half-cycles per AC cycle, the total switching events per cycle are four turn-on and four turn-off transitions. The corresponding switching-loss power is:
\begin{equation}
P_{\text{sw}}\!=\!f_{\text{sw}}\!\left(4E_{\text{on}}\!+\!4E_{\text{off}}\right)\!=\!\frac{2\sqrt{2}}{\pi} f_{\text{sw}} V_{\text{DC}} |\mathbf{I}_{\text{AC}}|\!\left(t_{\text{on}}\!+\!t_{\text{off}}\right)
\label{eq:P_sw_bridge}
\end{equation}

\noindent where $t_{\text{on}}$ and $t_{\text{off}}$ are the on and off transition durations.

% \paragraph{Diode Reverse Recovery Loss}
\noindent \textit{b) Diode Reverse Recovery Loss:} Reverse recovery loss occurs when diodes switch from forward conduction to reverse blocking. During this transition, stored charge is released as a reverse current pulse. In unipolar SPWM, only two diodes undergo this transition per cycle. We define the effective recovery duration as $t_{\text{Doff}} = Q_{\text{RR}} / I_{\text{TEST}}$ \cite{scapino2003transformer}. Assuming a constant reverse voltage of $V_{\text{DC}}/2$ and sinusoidal current \eqref{AC_current_inst}, the energy loss per event is
\begin{equation}
E_{\text{rr}} = \frac{V_{\text{DC}}}{2}  \frac{2\sqrt{2}}{\pi} |\mathbf{I}_{\text{AC}}|  t_{\text{Doff}}
\end{equation}

\noindent with two diodes undergoing reverse recovery events per switching cycle, the total reverse recovery power loss is:
\begin{equation}
P_{\text{rr}} = 2 f_{\text{sw}} E_{\text{rr}} = \frac{2\sqrt{2}}{\pi} f_{\text{sw}} V_{\text{DC}} t_{\text{Doff}} |\mathbf{I}_{\text{AC}}|
\label{eq:P_rr_total}
\end{equation}

\noindent \textit{c) Total Switching Loss:} The total switching loss is the sum of transistor switching losses and diode reverse recovery losses ($P_{\text{sw,total}} = P_{\text{sw}} + P_{\text{rr}}$). We represent this as a controlled current source on the DC side: $I_{\text{SW}} = P_{\text{sw,total}} / V_{\text{DC}}$.
\begin{equation}
I_{\text{SW}} = \frac{2 \sqrt{2}}{\pi} f_{\text{sw}} (t_{\text{on}} + t_{\text{off}} + t_{\text{Doff}})\,  |\mathbf{I}_{\text{AC}}|
\end{equation}

\noindent To express switching loss in terms of DC current, we substitute the AC current magnitude using the power balance relation \eqref{eq:ssc_power_balance} and voltage relation \eqref{eq:AC_RMS_mag}:

\begin{equation}
|\mathbf{I}_{\text{AC}}| = \frac{V_{\text{DC}} I_{\text{DC}}}{|\mathbf{V}_{\text{AC}}| \cos\phi} 
= \frac{V_{\text{DC}} I_{\text{DC}}}{\frac{M V_{\text{DC}}}{\sqrt{2}} \cos\phi}
= \frac{\sqrt{2}}{M \cos\phi} I_{\text{DC}}
\end{equation}

\noindent Substituting into the expression for switching current yields:
\begin{equation}
I_{\text{SW}} = \frac{4 f_{\text{sw}}}{\pi M \cos\phi} (t_{\text{on}} + t_{\text{off}} + t_{\text{Doff}})\, I_{\text{DC}}
\label{eq:I_SW_AC}
\end{equation}

\noindent where the modulation–power factor product is defined as:
\begin{equation}
M \cos\phi = \left( M^R I_{\text{AC}}^R + M^I I_{\text{AC}}^I \right) |\mathbf{I}_{\text{AC}}|^{-1}
\end{equation}
\end{subequations}

\subsubsection{\SSC{} Conduction Losses}

\noindent Conduction losses in the \SSC{} result from on-state voltage drops across transistors and diodes. Each device is modeled by the linear relation $v = V_0 + R i$, where $V_0$ is the threshold voltage and $R$ is the on-state resistance \cite{mohan2003power}. The instantaneous conduction power is:
\begin{subequations}
\begin{equation}
p_c(i) = i (V_0 + R i)
\label{eq:ssc_pinst}
\end{equation}

\noindent Averaging this over a full electrical cycle yields the average conduction loss \cite{berringer1995semiconductor}:
\begin{equation}
P_C = \frac{1}{2\pi} \int_{0}^{2\pi} p_c\left(i\left(\theta\right)\right) \, d\theta = \overline{I} V_0 + I_{\text{rms}}^2 R
\label{eq:cond_loss_general}
\end{equation}

\noindent where $\overline{I}$ is the average value of conduction current and $I_{\text{RMS}}$ is the RMS conduction current and applies to both transistors and diodes.

% \paragraph{Switching Duty and Conduction Current}
% In unipolar SPWM, the gating signal for each transistor is determined by comparing a sinusoidal modulation reference with a triangular carrier. The reference is defined as a function of the voltage angle as $v_{\text{ref}}(\theta_V) = M \frac{V_{\text{DC}}}{2} \sin(\theta_V)$. A transistor conducts when the reference exceeds the carrier signal, yielding a normalized conduction interval
% $D(\theta_V) = \frac{1}{2}[1 + M \sin(\theta_V)]$

% To express this in terms of the current phasor, we substitute $\theta_V = \theta_I + \phi$, where $\theta_I$ is the current angle and $\phi$ is the power factor angle. The transistor and diode duty cycles become:
% \begin{align}
% D_T(\theta_I) &= \frac{1}{2} (1 + M \sin(\theta_I + \phi)) \\
% D_D(\theta_I) &= 1 - D_T(\theta_I) = \frac{1}{2} (1 - M \sin(\theta_I + \phi))
% \end{align}

% Using the phase current expression from \eqref{AC_current_inst}, the conduction current in each device is the product of the sinusoidal current and the corresponding duty cycle:
% \begin{equation}
% i_T(\theta_I) = i(\theta_I) D_T(\theta_I + \phi),  i_D(\theta_I) = i(\theta_I) D_D(\theta_I + \phi)
% \end{equation}

% \paragraph{Switching Duty and Conduction Current}

\noindent \textit{a) Switching Duty and Conduction Current:}
\noindent In unipolar SPWM, the gating signal is generated by comparing a sinusoidal reference with a high-frequency triangular carrier. The reference, aligned with the voltage phasor, is expressed in the current reference frame as:
\begin{equation}
v_{\text{ref}}(\theta) = M \frac{V_{\text{DC}}}{2} \sin(\theta +\theta_v) = M \frac{V_{\text{DC}}}{2} \sin(\theta + \theta_i + \phi)
\end{equation}

\noindent where 
%$\theta = \omega t$, 
$\theta_i$ is the current phase angle of the current
%in the stationary (absolute) frame, 
$\theta_v$ represents the voltage phase angle, 
%in this same rotating frame, 
and $\phi = \theta_v - \theta_i$ is the phase difference between voltage and current. 
Conduction occurs when the reference waveform exceeds the carrier, yielding a duty ratio proportional to $\sin(\theta + \theta_v)$. The resulting transistor and diode duty cycles are:
\begin{align}
D_T(\theta) &
% = \frac{1}{2} \left(1 + M \sin(\theta + \theta_v)\right)
=\tfrac{1}{2} \left(1 + M \sin(\theta + \theta_i + \phi)\right)
\\
D_D(\theta) &= 1 - D_T(\theta) = \tfrac{1}{2} \left(1 - M \sin(\theta + \theta_i + \phi)\right)
\end{align}

\noindent We compute the average and squared conduction currents by integrating over the ON interval of each switching cycle with period $T_s$. Assuming the current $i(\theta)$ remains constant over $T_s$, the duty-weighted expressions for switch $x \in \{T, D\}$ become:

\begin{align}
i_x(\theta) &= \frac{1}{T_s} \int_0^{D_x(\theta) T_s} i(\theta)\,dt = i(\theta) D_x(\theta) \\
l_x(\theta) &= \frac{1}{T_s} \int_0^{D_x(\theta) T_s} i^2(\theta)\,dt = i^2(\theta) D_x(\theta)
\end{align}

\noindent Here, $i_x(\theta)$ represents the duty-scaled conduction current, while $l_x(\theta)$ denotes the duty-weighted squared current. 

% \paragraph{Conduction Current Symmetry and Averaging}

\noindent \textit{b) Conduction Current Symmetry and Averaging:} Each transistor and diode conducts over complementary half-cycles defined by the unipolar SPWM duty cycles. Due to symmetry in both the current waveform and the modulation logic, all transistors and all diodes exhibit identical conduction characteristics. We evaluate their average and RMS currents over a full conduction interval using $i_T(\theta)$ and $i_D(\theta)$, integrating over $[-\theta_i, -\theta_i + \pi]$ and normalizing by $2\pi$. The average currents are:
\begin{align}
\overline{I_T} &= \frac{1}{2\pi} \!\! \int_{-\theta_i}^{\pi -\theta_i} \!\! i_T(\theta)\, d\theta = \frac{\sqrt{2} I_{\text{AC}}}{8\pi}(4+\pi M\cos\phi) \\
\overline{I_D} &= \frac{1}{2\pi} \!\! \int_{-\theta_i}^{\pi-\theta_i} \!\! i_D(\theta)\, d\theta = \frac{\sqrt{2} I_{\text{AC}}}{8\pi}(4-\pi M\cos\phi)
\end{align}

% \noindent and the corresponding RMS currents are:
% \begin{align}
% I_{T,\text{rms}} &= \sqrt{\frac{1}{2\pi} \!\! \int_{-\theta_i}^{\pi-\theta_i} \!\! l_T(\theta)\, d\theta} = \frac{|\mathbf{I}_{\text{AC}}|}{6\sqrt{\pi}} \sqrt{9\pi + 24 M\cos\phi} \\
% I_{D,\text{rms}} &= \sqrt{\frac{1}{2\pi} \!\! \int_{-\theta_i}^{\pi-\theta_i} \!\! l_D(\theta)\, d\theta} = \frac{|\mathbf{I}_{\text{AC}}|}{6\sqrt{\pi}} \sqrt{9\pi - 24 M\cos\phi}
% \end{align}

\noindent and the corresponding RMS currents are:
{\relsize{-0.5} % Reduces the size by 5%
\begin{align}
I_{T,\text{rms}} &= \sqrt{\frac{1}{2\pi} \!\! \int_{-\theta_i}^{\pi-\theta_i} \!\! l_T(\theta)\, d\theta} = \frac{|\mathbf{I}_{\text{AC}}|}{6\sqrt{\pi}} \sqrt{9\pi + 24 M\cos\phi} \\
I_{D,\text{rms}} &= \sqrt{\frac{1}{2\pi} \!\! \int_{-\theta_i}^{\pi-\theta_i} \!\! l_D(\theta)\, d\theta} = \frac{|\mathbf{I}_{\text{AC}}|}{6\sqrt{\pi}} \sqrt{9\pi - 24 M\cos\phi}
\end{align}
}

\noindent Because $0 \le M \le 1$ and $\cos\phi \in [-1,1]$, the square root argument $9\pi \pm 24 M \cos\phi$ remains strictly positive for all feasible operating points. Since $M\cos\phi$ may be negative in certain inverter operating quadrants, we apply the absolute value to ensure that transistors and diodes receive symmetric average and RMS current assignments across all power flow conditions.

% Because $0 \le M \le 1$ and $\cos\phi \in [-1,1]$, the square root argument $9\pi \pm 24 M \cos\phi$ remains strictly positive for all feasible operating points.

% \noindent Since $M \cos\phi$ can be negative in certain inverter operating quadrants, we apply the absolute value to ensure that transistors and diodes receive equal average and RMS current assignments under all power flow conditions.
\noindent \textit{c) Total conduction Loss:} The total conduction loss in the \SSC{}, comprising four transistors and four diodes, is:
\begin{equation}
\begin{aligned}
P_{\text{C}} &= 
\frac{\sqrt{2}|I_{\text{AC}}|}{2\pi} \!
\left[\! V_T \!\left(\! \pi|M\cos\phi|\!+\!4 \!\right) 
\!+\! V_D \!\left(\! 4\!-\!\pi|M\cos\phi| \!\right)\! \right] \\
&\quad + \frac{|I_{\text{AC}}|^2}{3\pi} \!
\left[\! R_T \!\left(\! 8|M\cos\phi|\!+\!3\pi \!\right) 
\!+\! R_D \!\left(\! 3\pi\!-\!8|M\cos\phi| \!\right)\! \right]
\end{aligned}
\label{eq:total_conduction_loss}
\end{equation}

 We model conduction loss using a complex voltage source with real and imaginary components, $V_C^R$ and $V_C^I$, placed in series with the inverter output. The power dissipated by this source is expressed as:
\begin{equation}
P_C = V_C^R I_{\text{AC}}^R + V_C^I I_{\text{AC}}^I
\end{equation}

\noindent because conduction loss is strictly real, the voltage source will not introduce reactive power, leading to the constraint:
\begin{equation}
Q_C = V_C^I I_{\text{AC}}^R - V_C^R I_{\text{AC}}^I = 0
\end{equation}

\noindent this constraint ensures that the conduction voltage aligns with the current phasor, and this yields:
\begin{equation}
\begin{aligned}
V_C^R = \frac{P_C  I_{\text{AC}}^R}{|I_{\text{AC}}|^2}, \quad
V_C^I = \frac{P_C  I_{\text{AC}}^I}{|I_{\text{AC}}|^2}
\end{aligned}
\end{equation}
\end{subequations}

\begin{figure}[htpb]
    \centering
    \includegraphics[width=0.8\linewidth]{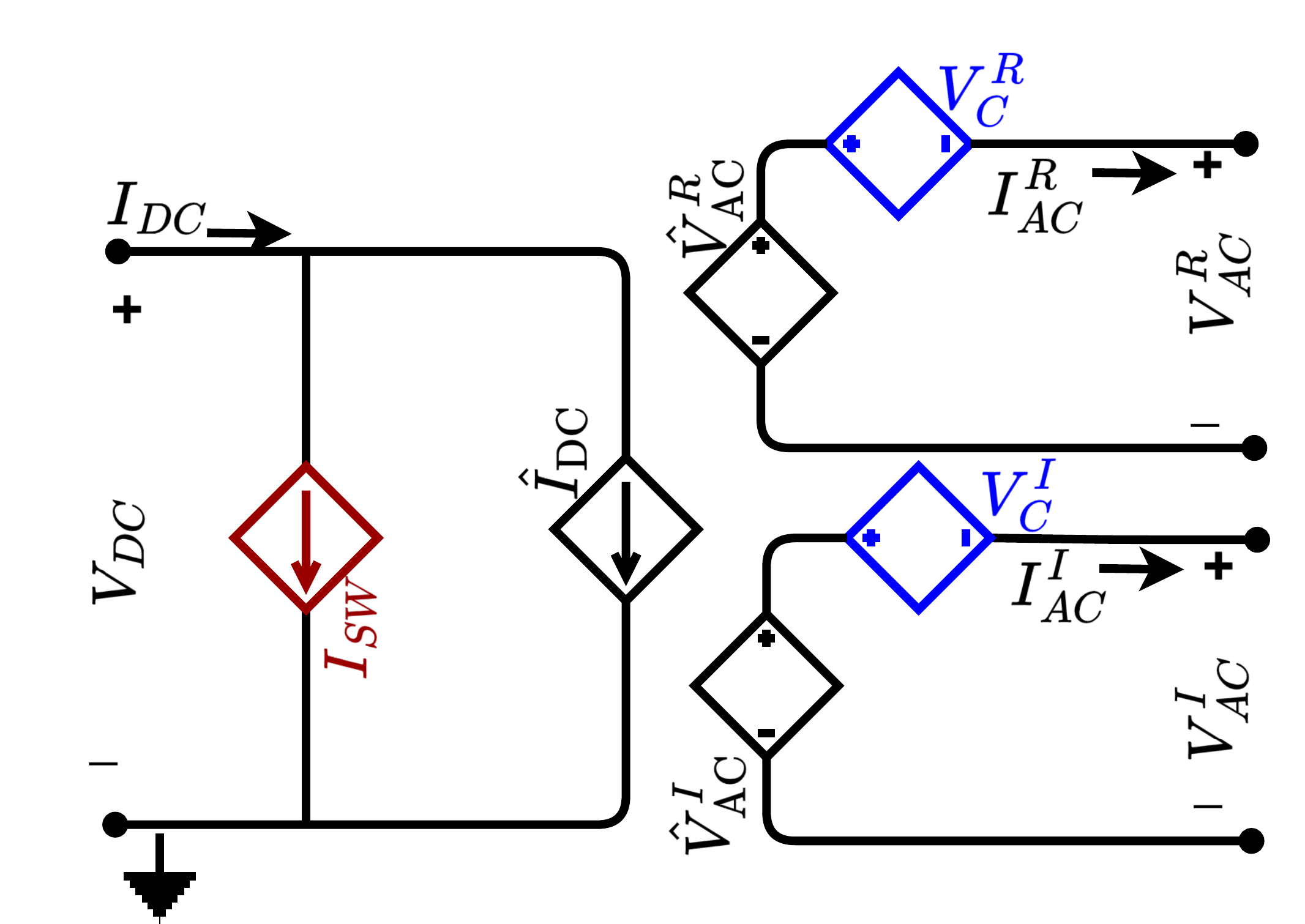}
    \caption{Equivalent non-ideal \SSC{} model with switching loss as controlled current source and conduction loss as controlled voltage source.}
    \label{fig:second_stage_converter_equiv}
\end{figure}

Fig. \ref{fig:second_stage_converter_equiv} shows the non-ideal \SSC{} circuit model. Switching and conduction losses are represented using controlled sources.

\subsection{AC Filter Model and Grid Interface}  
\label{subsec:AC_filter_model}  

\begin{figure*}[htpb]
    \centering
    \includegraphics[width=0.9\linewidth]{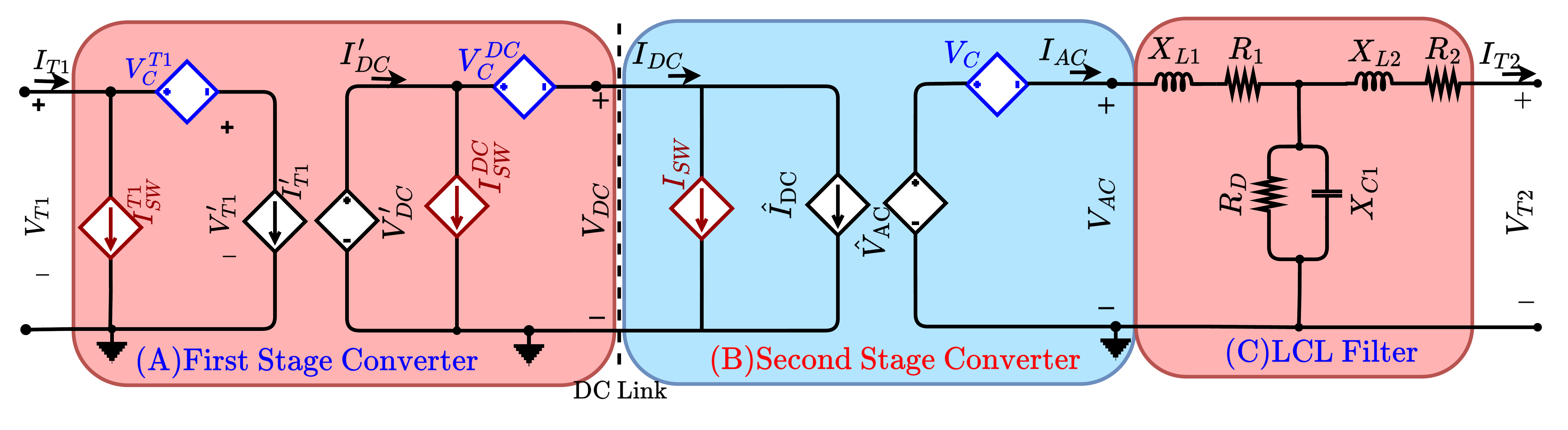}
    % \caption{Complete equivalent circuit model of \InverterShort{} comprising (A) \FSC{}, (B) \SSC{}, and (C) AC filter with DC link in between the \FSC{} and \SSC{}. A controlled current source connects with the secondary of \SSC{} depending on the choice of control mode in Section \ref{sec:Control}.}
    \caption{Complete equivalent circuit model of \InverterShort{}, comprising (A) \FSC{}, (B) \SSC{}, and (C) an AC LCL filter, with the DC link coupling the \FSC{} and \SSC{}. The model connects the DC side (\TOne{}) and AC side (\TTwo{}) through the unified algebraic functions \eqref{eq:fsc_phys_sum}--\eqref{eq:pcc_phys_sum}, which enforce power balance, capture converter losses, and ensure smooth consistency between electrical states and control variables across all operating modes mentioned in Section \ref{sec:Control}.}
    \label{fig:inverter_system_architecture}
\end{figure*}

\noindent The output of the \SSC{} contains high-frequency harmonics due to unipolar SPWM switching. 
To attenuate these and comply with grid quality standards, the general practice is to include a $LCL$-$P$-$R$ filter\cite{Yao2016Design}. 
The topology consists of a converter-side (terminal $AC$) inductor $L_1$, a grid/load-side (terminal $\TTwo{}$) inductor $L_2$, and a damping branch comprising a capacitor $C$ in series with a resistor $R_d$.
We denote these in steady-state with their equivalent impedances denoted as $X_{L1}$, $X_C$, and $X_{L2}$ in block (C) of Fig. \ref{fig:inverter_system_architecture}. Parasitic resistances $R_1$ and $R_2$ model conduction losses while the damping resistor $R_d$ suppresses resonance near the filter’s natural frequency.

%Applying KCL and KVL yields the steady-state phasor equations in rectangular form:
%\begin{subequations}
%\begin{align}
%I_{\text{T2}}^R &= I_{\text{AC}}^R - R_d^{-1} (V_{\text{AC}}^R + \omega L_1 I_{\text{AC}}^I - R_1 I_{\text{AC}}^R) \notag \\
%&\quad + \omega C (V_{\text{AC}}^I - \omega L_1 I_{\text{AC}}^R - R_1 I_{\text{AC}}^I) \\
%I_{\text{T2}}^I &= I_{\text{AC}}^I - R_d^{-1} (V_{\text{AC}}^I - \omega L_1 I_{\text{AC}}^R - R_1 I_{\text{AC}}^I) \notag \\
%&\quad - \omega C (V_{\text{AC}}^R - R_1 I_{\text{AC}}^R + \omega L_1 I_{\text{AC}}^I) \\
%V_{\text{T2}}^R &= V_{\text{AC}}^R - R_1 I_{\text{AC}}^R - R_2 I_{\text{T2}}^R - \omega (L_1 I_{\text{AC}}^I + L_2 I_{\text{T2}}^I) \\
%V_{\text{T2}}^I &= V_{\text{AC}}^I + R_1 I_{\text{AC}}^I + R_2 I_{\text{T2}}^I - \omega (L_1 I_{\text{AC}}^R + L_2 I_{\text{T2}}^R)
%\end{align}
%\end{subequations}

%\subsection{Full Model: System Integration and Power Interface}
The full equivalent circuit for \InverterShort{} model by connecting the \FSC{}, \SSC{}, and LCL filter is shown in Fig. \ref{fig:inverter_system_architecture}.
It shows the integrated structure between terminal $T_1$ (DER side) and terminal $T_2$ (load/grid side), supporting both inversion and rectification modes. 
Section \ref{sec:Control} discusses how the inverter $T_2$ terminal interfaces with the grid based on different control modes.

%The inverter exchanges real and reactive power with the grid at terminal $T_2$ as:
%\begin{subequations}
%\begin{align}
%P &= V_{T2}^{R} I_{T2}^{R} + V_{T2}^{I} I_{T2}^{I} \label{eq:active_power} \\
%Q &= V_{T2}^{I} I_{T2}^{R} - V_{T2}^{R} I_{T2}^{I} \label{eq:reactive_power}
%\end{align}
%\end{subequations}

% \begin{figure*}[htpb]
%     \centering
%     \includesvg[width=0.9\linewidth]{images/inverter_full.drawio.svg}
%     \caption{Complete equivalent circuit model of  \InverterShort{} comprising (A) \FSC{}, (B) \SSC{}, and (C) AC filter with DC link in between the \FSC{} and \SSC{}. A controlled current source connects with the secondary of \SSC{} depending on the choice of control mode in Section \ref{sec:Control}.}
% \label{fig:inverter_system_architecture}
% \end{figure*}

\section{Control Strategies}\label{sec:Control}

% \noindent Inverters—such as the \InverterShort{} modeled in this work—act as essential interfaces between DC-based distributed energy resources and the AC grid, influencing voltage and power flow at both terminals. 
% Inverter's internal physics and its control functions determines its behavior.
% %This behavior is determined by the inverter’s internal physics and its control functions. 
% We represent the internal physics and losses as circuit equivalents in Section \ref {formulation}. 
% Control functions are embedded to regulate the inverter operation based on conditions at the two end terminals. 
% In this work, we categorize the control into active power control and reactive power control.

% --------------------------
% We model the inverter control with controlled current sources connected to the secondary of second-stage converter, without loss of generality:

\noindent We model the inverter control with controlled current sources $(I_{T2})$ connected to the terminal \TTwo{} of the \InverterShort{} and serving as agrid/load interface. We define the controlled source expressions as a function of $P_{ctrl}$ and $Q_{ctrl}$:
\begin{subequations}
\begin{align}
I_{T2}^{R} = \frac{P_{ctrl}\,V_{T2}^{R} + Q_{ctrl}\,V_{T2}^{I}}{(V_{T2}^{R})^{2} + (V_{T2}^{I})^{2}},
I_{T2}^{I} = \frac{P_{ctrl}\,V_{T2}^{I} - Q_{ctrl}\,V_{T2}^{R}}{(V_{T2}^{R})^{2} + (V_{T2}^{I})^{2}}\label{eq:ctrl_current}
\end{align}

\noindent We discuss two categories: active power control and reactive power control that define expressions for $P_{\text{ctrl}}$ and $Q_{\text{ctrl}}$.

% \subsection{Active Power Control}  
% \noindent Active power control regulates real power exchange between the DC and AC sides. 
% We consider two modes: demand-driven, where power setpoints are specified, and supply-driven, where output follows available DC energy. 
% In the latter, the inverter adapts to source conditions, such as SOC for batteries or maximum power point tracking (MPPT) for PV systems.
% We model all control via current-controlled sources connected to the grid.

% \subsection{Active Power Control}  
% \noindent Active power control regulates real power exchange between the DC and AC sides. We consider two modes: demand-driven, such as constant $P$ control, where power setpoints are specified, and supply-driven, where output follows available DC energy. In the latter, the inverter adapts to source conditions, such as SOC for batteries or maximum power point tracking (MPPT) for PV systems. In both modes, we implement this control as constant current sources.

\subsection{Active Power Control}  
\noindent
We consider two control modes: demand-driven constant $P$ control and supply-driven maximum power point tracking (MPPT).
%In both cases, we model the control using constant current sources.
%\noindent \paragraph{Constant $P$ Control}: 

\noindent \textit{a) Constant P Control:} In constant $P$ control, the inverter injects a specified active power $P_{\text{set}}$ into the grid:
\begin{align}
    P_{\text{ctrl}} = P_{\text{set}} \label{eq:constant_P_control}
\end{align}

% \begin{equations}
% I^{\text{ctrl}_R} = \frac{P_{set} V^R_{AC}}{{V^R_{AC}}^2 + {V^I_{AC}}^2}//
% I^{\text{ctrl}_I} = \frac{P_{set} V^I_{AC}}{{V^R_{AC}}^2 + {V^I_{AC}}^2}
% \end{equations}

%This demand-driven mode computes the required current from the terminal voltage and applies it as a constant current source.

\noindent \textit{b) MPPT Control:}  MPPT maximizes power extraction from PV source $P_{\text{MP}}$ by adjusting its operating voltage $V_{\text{T1}}$ to match the maximum power point voltage $V_{\text{MP}}$. This condition is satisfied when the derivative of power with respect to voltage vanishes \cite{jereminov2016improving}:
\begin{align}
\frac{dP_{\text{T1}}}{dV_{\text{T1}}} = I_{\text{T1}} + V_{\text{T1}} \frac{dI_{\text{T1}}}{dV_{\text{T1}}} = 0
\label{eq:mpp_condition}
\end{align}

% In practice, the controller compares $I_{\text{T1}} / V_{\text{T1}}$ which is $ G_{\text{inst}}$ and $dI_{\text{T1}} /dV_{\text{T1}}$ (approximated using $\Delta I_{\text{T1}} / \Delta V_{\text{T1}}$) which is $G_{\text{inc}}$  to determine if the PV is operating above or below the MPP and adjusts the operating voltage accordingly.

\noindent Using the I–V relationship in \eqref{eq:pv_iv_final} and applying the MPP condition in \eqref{eq:mpp_condition}, we derive a closed-form expression for the MPP operating point. Let $V_D = V_{\text{MP}} + I_{\text{MP}} R_s$, $\chi = \exp(V_D / V_{\text{th}})$, and $V_{\text{th}} = n N_s V_t$. The corresponding expressions are:
\begin{align}
I_{\text{MP}} &= I_{\text{ph}} - I_0(\chi - 1) - \frac{V_D}{R_{\text{sh}}}
\label{eq:mpp_iv_chi} \\
I_{\text{MP}} &= \frac{V_{\text{MP}} \left( I_0 R_{\text{sh}} \chi + V_{\text{th}} \right)}{I_0 R_s R_{\text{sh}} \chi + V_{\text{th}}(R_s + R_{\text{sh}})}
\label{eq:mpp_closed_form}
\end{align}

\noindent Solving \eqref{eq:mpp_iv_chi} and \eqref{eq:mpp_closed_form} simultaneously yields the MPP, which gives the PV power $P_{MP}$ and is provided to the AC side as the active-power control:
% \begin{align}
%     P_{\text{ctrl}} = P_{\text{MP}} - P_{\text{loss,inv}} = V_{\text{MP}} \cdot I_{\text{MP}} - P_{\text{loss,inv}}
% \end{align}
\begin{align}
P_{ctrl}
&= P_{MP} - P_{loss,inv} 
   = V_{MP} I_{MP} - P_{loss,inv} \notag\\
&= V_{T1} I_{T1}\big|_{MPP} - P_{loss,inv}
\end{align}
\label{eq:MPPT}
\end{subequations}
where $P_{\text{loss,inv}}$ denotes the total inverter loss, including conduction and switching losses across both converter stages and losses in the filter.

% \begin{figure}[htbp]
%     \centering
%     \includesvg[width=1\linewidth]{images/MPP_VV_combined.svg}
%     \caption{Control characteristics of the TSBI: (A) MPPT on I–V and P–V curves; (B) Illustration of piecewise and smooth Volt–VAR control curves.}
%     \label{fig:control_strategies}
% \end{figure}
\begin{figure}[htbp]
    \centering
    \includegraphics[width=1\linewidth]{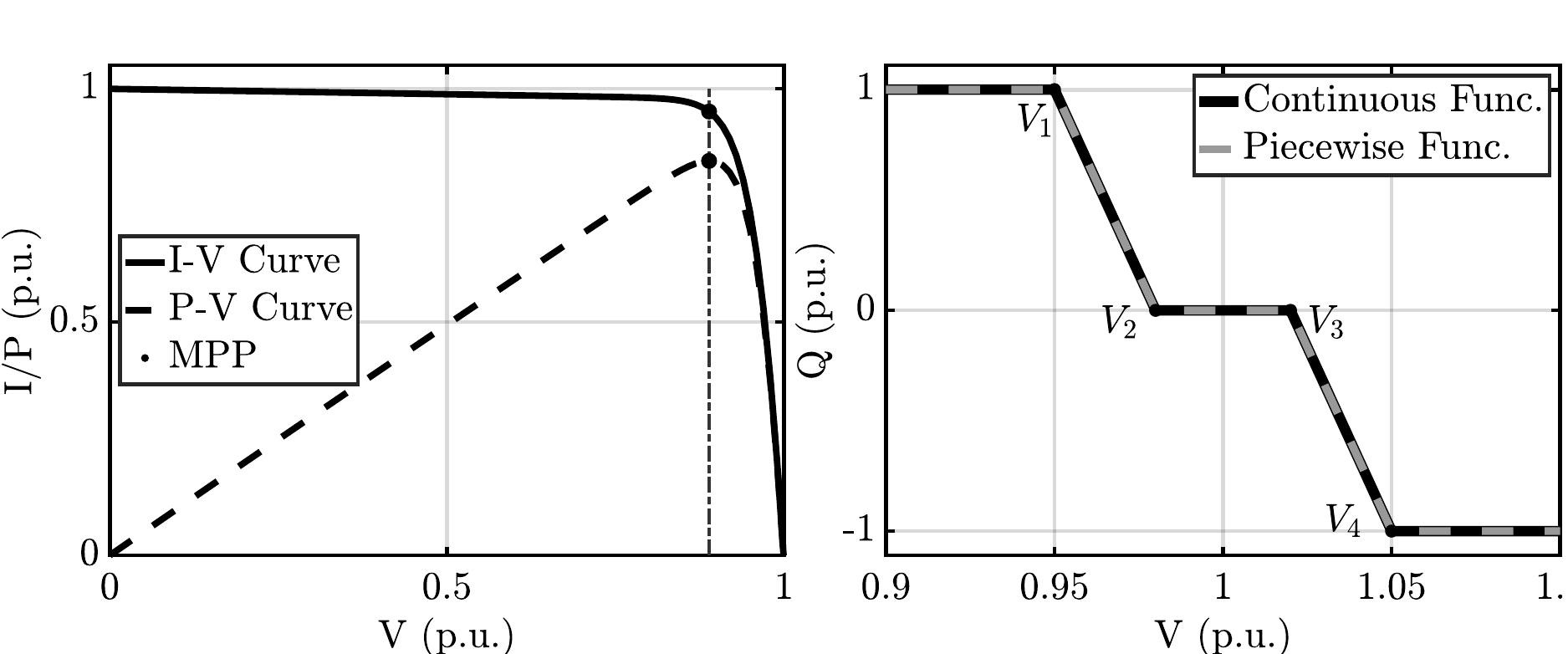}
    \caption{Control characteristics of the TSBI: (A) MPPT on I–V and P–V curves; (B) Illustration of piecewise and smooth Volt–VAR control curves.}
    \label{fig:control_strategies}
\end{figure}

\subsection{Reactive Power Control}  
\noindent We discuss three strategies: constant $Q$, constant power factor, and Volt–VAR control. 
We show how these strategies contribute to $Q_{\text{ctrl}}$ value in \eqref{eq:ctrl_current}.

\noindent \textit{a) Constant Q Control:} In constant $Q$ control, the inverter maintains a fixed reactive power output at terminal $T_2$, independent of voltage or load variations:
\begin{subequations}
\begin{align}
    Q_{\text{ctrl}} = Q_{\text{set}}
\end{align}
%Given $Q_{\text{set}}$ and the terminal voltage $V_{T2}$, the required reactive current is computed from Equation \eqref{eq:reactive_power}, and the inverter adjusts switching to sustain this output. 
% This strategy is simple and aligns with schemes like the Commelec framework \cite{bernstein2015composable}, where reactive setpoints are dispatched centrally. This control can be used in both active power control modes: for supply-driven cases, $P = f(I_{T1}, V_{T1})$ and $Q = Q_{\text{set}}$; for demand-driven cases, the inverter adjusts its internal states such that $I_{T1}, V_{T1} = f(P_{\text{set}}, Q_{\text{set}})$.

\noindent \textit{b) Constant PF Control:} In constant power factor control, the inverter sets reactive power $Q_{\text{ctrl}}$ to maintain a fixed power factor (PF) relative to the active power $P_{\text{ctrl}}$ at the point of interconnection. The relationship is given by:
\begin{equation}
Q_{\text{ctrl}}\ \cdot \text{PF} = P_{\text{ctrl}} \sqrt{1 - \text{PF}^2}
\end{equation}

% \noindent This approach can be seen in common standards like California Rule 21 \cite{johnson2014draft}. This control can be used in both active power control modes: supply-driven with $Q = f(P, \text{PF})$, and demand-driven with $I_{T1}, V_{T1}, Q = f(P_{\text{set}}, \text{PF})$.

\noindent \textit{c) Volt-VAR Control:} Volt–VAR control regulates terminal voltage by adjusting the inverter’s reactive output $Q_{\text{ctrl}}$ according to a piecewise curve:

\begin{equation}
Q_{\text{ctrl}}(V) = \begin{cases}
\overline{Q}, & V \le V_1 \\
\overline{Q}(V_2{-}V)/(V_2{-}V_1), & V_1 < V < V_2 \\
\underline{Q}(V{-}V_3)/(V_4{-}V_3), & V_3 < V < V_4 \\
0, & V_2 \le V \le V_3\\
\underline{Q}, & V \ge V_4
\end{cases}
\label{eq:piecewise}
\end{equation}

\noindent To solve with Newton-based methods, we approximate the piecewise curve using a smooth indicator function $f(x;a,b)$ \cite{amhraoui2022smoothing}:
 
\begin{equation}
f(x;a,b)=\frac{x-a+|x-a|}{2}-\frac{x-b+|x-b|}{2}a
\label{eq:piecewise_indicator}
\end{equation}
% \noindent which extracts the positive part of $x$ over $[a, b]$. Some solvers handle absolute value terms $|x|$ natively; this can be replaced with $|x| = \sqrt{x^2}$ just as in Section \ref{subsec:sign_function_modeling} without requiring smoothing, as we do not have to worry about dividing by zero for solvers that cannot.

\noindent This yields a differentiable Volt–VAR expression:
\begin{equation}
\begin{gathered}
Q_{\text{ctrl}}(V) =\; \overline{Q} - \\
\frac{\overline{Q}}{V_2-V_1} \left[\frac{V-V_1+|V-V_1|}{2}-\frac{V-V_2+|V-V_2|}{2}\right] +\\
\frac{\underline{Q}}{V_4-V_3} \left[\frac{V-V_3+|V-V_3|}{2}-\frac{V-V_4+|V-V_4|}{2}\right]
\end{gathered}
\label{eq:QV_single_function}
\end{equation}
\end{subequations}

\begin{added}{Reviewer 2 Comment 5}
\noindent We applied the $\epsilon$-regularization from \textit{Remark 1} so that all absolute-value terms in \eqref{eq:QV_single_function} are expressed as $|x|_\epsilon = \sqrt{x^{2}+\epsilon}$, $\epsilon>0$. This guarantees that $Q_{\text{ctrl}}(V)$ remains smooth and continuously differentiable across all voltage regions. This smooth form preserves the original control shape (see Fig. \ref{fig:control_strategies}B) and ensures convergence in nonlinear solvers. 
\end{added}

\subsection{TSBI Steady-State Model and Control}
\label{subsec:tsbi_model_io_pcc}

\begin{added}{Editor Comment 2, Reviewer 2 Comment 1, and Reviewer 3 Comment 4}

\noindent 
We use the equivalent circuit approach to model DC components such as PV and battery (see Section \ref{preliminaries}), the three-phase distribution grid feeder physics (see Section \ref{preliminaries}), as well as the bidirectional inverter model, \InverterShort{} (see Section \ref{formulation}).

The algebraic equation set representing the overall \InverterShort{} physics is given in \eqref{eq:fsc_phys_sum}--\eqref{eq:pcc_phys_sum}. 
The \InverterShort{} connects to the DC component (PV or battery) at the DC terminal (\TOne{}), while the AC terminal (\TTwo{}) couples with the distribution feeder through controlled current sources with control variables $P_{ctrl}$ and $Q_{ctrl}$ (see \eqref{eq:ctrl_current}). 
These AC-side currents appear directly in the feeder’s real and imaginary KCL relations \eqref{eq:kcl_real}–\eqref{eq:kcl_imag}, so the inverter states and network voltages are solved jointly. 
The control logic in Section \ref{sec:Control} specifies the relationships for $P_{ctrl}$ and $Q_{ctrl}$.  
Equations \eqref{eq:fsc_phys_sum}–\eqref{eq:lcl_phys_sum} encode the internal KCL/KVL relationships of the \FSC{}, \SSC{}, and the filter.

For each inverter, the new set of unknown electrical states includes terminal 1, terminal 2, and DC-link currents 
$(V_{T1}, I_{T1}, I_{DC}, V_{T2}, I_{T2})$.
We also introduce states for duty-cycle $(D)$ and modulation variables $(M, \theta_V)$.
The set of fixed parameters includes inverter resistances, capacitances, inductors, DC-link voltage, and switching frequency
$(\bm{\Psi}=\{R_T,R_D,R_L,L_1,L_2,C,R_d,V_{DC},f_{sw}\})$.
\begin{subequations}\label{eq:inv_full_model}
\begin{align}
\mathbf{f}_{FSC}\!\big(V_{T1},\!I_{T1},\!D,\!V_{DC},\!I_{DC},\!\bm{\Psi}\big)&=0 \label{eq:fsc_phys_sum}\\
\mathbf{f}_{SSC}\!\big(V_{AC}^{R},\!V_{AC}^{I},\!I_{AC}^{R},\!I_{AC}^{I},\!\{M,\!\theta_V\},\!V_{DC},\!I_{DC},\!\bm{\Psi}\big)&=0 \label{eq:ssc_phys_sum}\\
\mathbf{f}_{LCL}\!\big(V_{AC}^{R},\!V_{AC}^{I},\!I_{AC}^{R},\!I_{AC}^{I},\!V_{T2}^{R},\!V_{T2}^{I},\!I_{T2}^{R},\!I_{T2}^{I},\!\bm{\Psi}\big)&=0 \label{eq:lcl_phys_sum}\\
\mathbf{f}_{CTRL}\!\big(P_{ctrl},\!Q_{ctrl},\!V_{T2}^{R},\!V_{T2}^{I},\!I_{T2}^{R},\!I_{T2}^{I}\big)&=0 \label{eq:pcc_phys_sum}
\end{align}
\end{subequations}

\noindent
In \eqref{eq:inv_full_model}, $\mathbf{f}_{\text{FSC}}$ represents the first-stage converter current balance with losses (Sec. \ref{subsec:first_stage_converter_model}, \eqref{eq:dc_voltage_relation}–\eqref{eq:VC_DC}); 
$\mathbf{f}_{\text{SSC}}$ represents the SSC modulation/power balance with switching and conduction losses (Sec. \ref{subsec:second_stage_converter_model}, \eqref{eq:ssc_power_balance}–\eqref{eq:total_conduction_loss}); 
$\mathbf{f}_{\text{LCL}}$ encodes the LCL phasor relationships (Sec. \ref{subsec:AC_filter_model}); 
$\mathbf{f}_{\text{CTRL}}$ implements the control strategy via $P_{ctrl},Q_{ctrl}$ (Sec. \ref{sec:Control}, \eqref{eq:ctrl_current}).

\end{added}

\begin{added}{Reviewer 3 Comment 7}
The \InverterShort{} preserves generality as one unified set of algebraic equations governs all operating quadrants, power-flow directions, and control modes. Sign-aware smooth loss expressions eliminate mode switching, and the active and reactive power references $(P_{ctrl},Q_{ctrl})$ are incorporated through a single differentiable controlled current-source mapping. Consequently, the formulation applies directly to three-phase steady-state power-flow analysis and large-scale nonlinear optimization without loss of generality across different operating and control modes.
\end{added}

% \section{Model Validation}\label{model_validation}
% \input{validation}

% \section{Experimental Setup} \label{sectionIV}
% \input{Experimental_setup}

\section{Results}\label{Results}
\noindent We evaluate the \InverterShort{} model  along three key dimensions: 
(i) model accuracy, 
(ii) ability to model different controls without loss of generality,
(iii) scalability. We select all \InverterShort{} parameters based on manufacturer datasheets and prior studies. See Table \ref{tab:component_params} for details.

\renewcommand{\arraystretch}{0.9}
\begin{table}[htpb]
\centering
\caption{Inverter component parameters in \InverterShort{} model}
\begin{tabular}{@{}p{3cm} p{5.2cm}@{}}
\toprule
\textbf{Component} & \textbf{Parameters [Unit]} \\
\midrule
MOSFET (SPW47N60C3) \cite{InfineonSPW47N60C3} & 
$V_{T0} = 0.30$ V, $R_T = 25$ m$\Omega$, 
$t_d^{\text{on}} = 14$ ns, $t_r = 15$ ns, 
$t_d^{\text{off}} = 58$ ns, $t_f = 11$ ns \\
\midrule
Diode (MUR460) \cite{VishayMUR460} & 
$V_{D0} = 1.10$ V, $R_D = 50$ m$\Omega$, 
$t_{\text{rr}} = 75$ ns \\
\midrule
FSC Inductor (Würt) \cite{WurthInductor} & 
$R_L = 1.8$ m$\Omega$ \\
\midrule
Switching Frequencies & 
FSC = 50 kHz \cite{TILM5118}, SSC = 16 kHz \cite{SMASunnyBoy}\\
\midrule
LCL Filter \cite{reznik2013lcl} & 
$L_1 = 2.23$ mH, $L_2 = 0.045$ mH, $C_f = 15\,\mu$F, 
$R_D = 0.55\,\Omega$, 
$R_1, R_2 < 5$ m$\Omega$ \\
\midrule
Ratings (Powerwall 3) \cite{TeslaPowerwall3} & 
$S = [5.8, 7.6, 10,11.5]$ kVA \\
% \midrule
% PV Module (LG400) \cite{LGLG400N2WV5} & $V_{\text{MP}} = 40.6$ V, 
% $I_{\text{MP}} = 9.86$ A \\
\bottomrule
\end{tabular}
\label{tab:component_params}
\end{table}

\noindent Additionally, for the battery model we use the Tesla Powerwall~3 energy capacity $E_{cap} = 13.5$\,kWh \cite{TeslaPowerwall3}. 
We assume a nominal open-circuit voltage of $V_{OC,nom} = 50$\,V and an internal resistance of $R_{int} = 36$\,m$\Omega$ for use in the zeroth-order battery equivalent circuit. 
For the PV model, we use the LG400 module with $V_{MP} = 40.6$\,V and $I_{MP} = 9.86$\,A \cite{LGLG400N2WV5}.

\subsection{Validation Against Time-Domain Detailed Model}
\label{subsec:simulink_validation}
\begin{added}{Editor Comment 4 and Reviewer 3 Comment 5}

\noindent
We validate the proposed \InverterShort{} against processed steady-state results from a time-domain switching benchmark implemented in Simulink–Simscape \cite{miller2009modeling} using the identical component parameters in Table \ref{tab:component_params}. 
The benchmark replicates the inverter topology, comprising an SPWM controller, transistors, diodes, and an $LCL$ filter that includes inductor parasitics and damping resistance. The carrier-based SPWM scheme uses a 16\,kHz triangular carrier and a 60\,Hz sinusoidal reference, while a constant DC source supplies $P=1440$\,W to the converter.
Simulations are performed using a fixed-step \texttt{ode4} (Runge–Kutta) solver with a time step of $1\,\mu\text{s}$. To enable comparison with our steady-state model, $60$\,Hz RMS and mean quantities are extracted from the time-domain simulation using measurement blocks. We report transistor and diode currents ($\overline{I}_T$, $I_{T,\text{rms}}$, $\overline{I}_D$, $I_{D,\text{rms}}$) and the average conduction drop ($V_{\text{cond}}$). 
The steady-state model closely matches the time-domain results, with current deviations of less than 2.7\% and voltage deviations of under 4.0\%. See Table \ref{tab:validation_results}. All validation data and files are available at \cite{badmus2024validation}.

\begin{table}[htbp]
\centering
\caption{Comparison of Steady-State Values and Errors Between TSBI and Simulink Benchmark}
\label{tab:validation_results}
\renewcommand{\arraystretch}{1.1}
\setlength{\tabcolsep}{5pt}
\begin{tabular}{lccccc}
\toprule
\textbf{Model} & $\overline{I}_T$ & $I_{T,\text{rms}}$ & $\overline{I}_D$ & $I_{D,\text{rms}}$ & $V_{\text{cond}}$ \\
\midrule
TSBI (steady state) & 4.4612 & 7.7975 & 0.8893 & 3.1363 & 1.4567 \\
Simulink (time domain) & 4.4310 & 7.8160 & 0.8662 & 3.0870 & 1.4010 \\
\midrule
\textbf{Error (\%)} & \textbf{0.68} & \textbf{0.24} & \textbf{2.67} & \textbf{1.60} & \textbf{3.97} \\
\bottomrule
\end{tabular}
\vspace{2mm}
\begin{minipage}{0.47\textwidth}
\footnotesize
\emph{Note:} Errors are computed as 
$\varepsilon(x)=\frac{|x_{\text{TSBI}}-x_{\text{Sim}}|}{|x_{\text{Sim}}|}\times 100\%$. 
Simulink values are $60$\,Hz steady-state averages and RMS extracted from time-domain switching simulations.
\end{minipage}
\end{table}

\end{added}

\begin{added}{Reviewer 1 Comment 3}

\subsection{Analysis of Efficiency on Model Accuracy}
\label{subsec:accuracy}

\noindent Next, we analyze the efficiency of the \InverterShort{} with varying input active and reactive power.  
Fig. \ref{fig:efficiency} presents the efficiency surface $\eta(P,Q)$ as a function of active and reactive power, showing the nonlinear variation with both $P$ and $Q$.

\begin{figure}[htpb]
    \centering
    \includegraphics[width=0.9\linewidth]{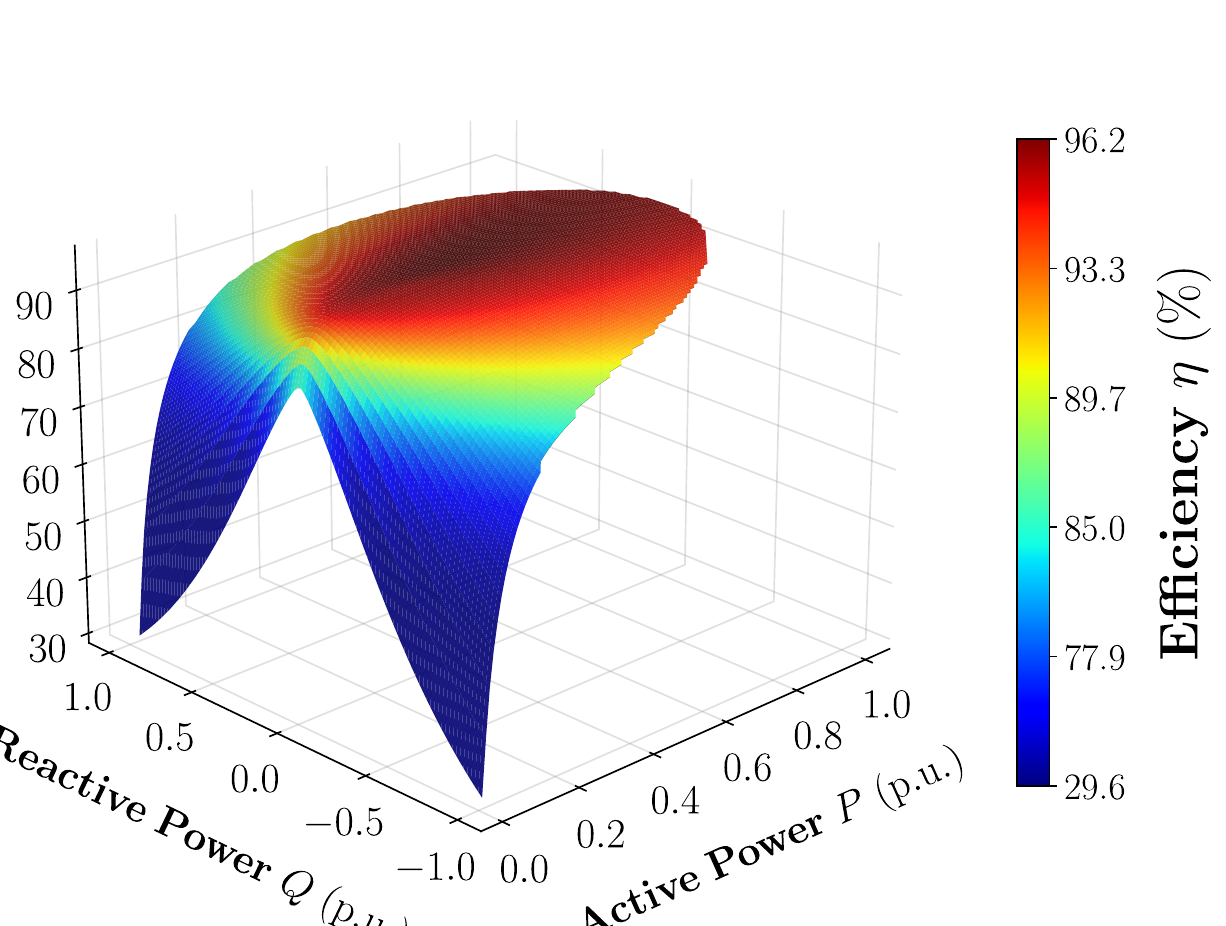}
    \caption{\InverterShort{} model efficiency as a function of active and reactive power, highlighting losses under varying $P$–$Q$ conditions.}
    \label{fig:efficiency}
\end{figure}

Inverter losses vary nonlinearly with output power, which itself depends quadratically on current magnitude.
Conduction power losses include a component that varies linearly with current and a resistive component that increases with the square of current \eqref{eq:VC_T1}–\eqref{eq:VC_DC}, \eqref{eq:cond_loss_general}. 
Switching power losses increase approximately linearly with current \eqref{eq:switching_current_t1_sgn}–\eqref{eq:switching_current_dc_sgn}, \eqref{eq:I_SW_AC}, whereas resistive power losses in the LCL filter grow with the square of the current. 
These dependencies produce the three characteristic regions observed in Fig. \ref{fig:efficiency}:

\begin{enumerate}[leftmargin=1em]
\item \noindent\textit{ Efficiency at low power:}
At light loading, where output current is small, linear-in-current losses from device voltage drops and switching transitions decrease more slowly than the quadratic growth of output power ($P_{out}\!\propto\! I_{rms}^{2}$), causing $P_{loss}/P_{out}$ to increase sharply and efficiency to drop.

\item \textit{Efficiency peak and roll-off:}
As the inverter transitions from light loading to higher current levels, output power increases quadratically with current, while linear conduction and switching losses grow more slowly. 
This causes the loss fraction $P_{loss}/P_{out}$ to decrease, allowing efficiency to rise toward its maximum. 
Beyond this point, the quadratic conduction losses in the semiconductor devices and passive components rise rapidly, while the linear current-dependent losses continue to add to the total power loss, causing $P_{loss}$ to grow faster than $P_{out}$ and resulting in the gradual efficiency roll-off after the maximum point. 

\item \textit{Dependence on reactive power:}
For a fixed active power $P$, variations in reactive power $Q$ modify the phase angle $\varphi$ between voltage and current. 
An increase in $|Q|$ therefore lowers the power factor $\cos\varphi$ and raises the current magnitude $I_{AC}=P/(V_{AC}\cos\varphi)$, which increases conduction and switching losses, reducing efficiency for both leading and lagging conditions. 
A slight local maximum occurs at small capacitive $Q$, where the leading current partially offsets the inductive reactive demand of the LCL filter.
\end{enumerate}

\end{added}

\subsection{Three-phase Power Flow on a Large Feeder}
\label{subsec:scalability}

\noindent We evaluate the scalability and control integration capabilities of the \InverterShort{} model on the large-scale r3\_12\_47\_3 distribution feeder \cite{schneider2014ieee}, which contains 16,239 nodes and 1,625 inverter-equipped load buses. Each inverter is rated at 10 kVA and dispatches 9 kW of active power (0.9 p.u.), reserving capacity for reactive support. We test three reactive control strategies: unity power factor (UPF), constant power factor (CPF), and Volt-VAR (Volt-VAR). For Volt-VAR, we adopt IEEE 1547-2018 Category A settings \cite{photovoltaics2018ieee}, which prescribe a piecewise curve with a maximum $|Q|$ of 0.25 p.u. To ensure fair comparison, we configure CPF to inject a fixed reactive power equal to the peak value used by Volt-VAR.

\begin{added}{Reviewer 1 Comment 4}
The three-phase power flow is formulated using the unbalanced AC ECF in Section \ref{sec:Grid_formulation}, expressed in rectangular I-V coordinates \eqref{eq:kcl_real}–\eqref{eq:kcl_imag}. 
The inverter equations \eqref{eq:fsc_phys_sum}–\eqref{eq:pcc_phys_sum} are solved implicitly with the network KCL constraints to enforce nodal balance. 
All cases use the interior-point solver IPOPT \cite{wachter2006implementation} (implemented as a feasibility problem with a constant objective), which exploits the continuous differentiability of the ECF–TSBI formulation to achieve efficient convergence. 
\end{added}

The formulation scales consistently across the three control modes. 
The UPF case results in 68,905 variables and constraints. 
For CPF, each inverter introduces two additional constraints: one to enforce a constant power factor and another to compute the apparent power $S$, thereby increasing the problem size by twice the number of inverters (1,625), which totals 72,155 variables and constraints. 
The Volt-VAR case adds three constraints per inverter: one to compute local voltage magnitude, one to implement the Volt-VAR curve, and one to compute $S$, resulting in 73,780 variables and constraints.
%No inequality constraints are required, as all control actions and limits are encoded within the physics-based formulation. 
We achieve solver convergence in all cases: UPF and CPF solved in under 2.5 seconds with eight iterations. Volt-VAR required 291 iterations and 62 seconds. 
These results confirm that the \InverterShort{} model enables accurate and scalable integration of inverter behavior in control-aware unbalanced power flow for large distribution networks.

% \begin{figure*}[htpb]
%     \centering
%     \includesvg[width=0.95\linewidth]{images/voltage_comparison.svg}
%     \caption{Steady-state voltage profiles across the feeder under three reactive control strategies (UPF, CPF, Volt-VAR) at a fixed active power injection per inverter.}
%     \label{fig:voltage_comparison}
% \end{figure*}
\begin{figure*}[htpb]
    \centering
    \includegraphics[width=0.9\linewidth]{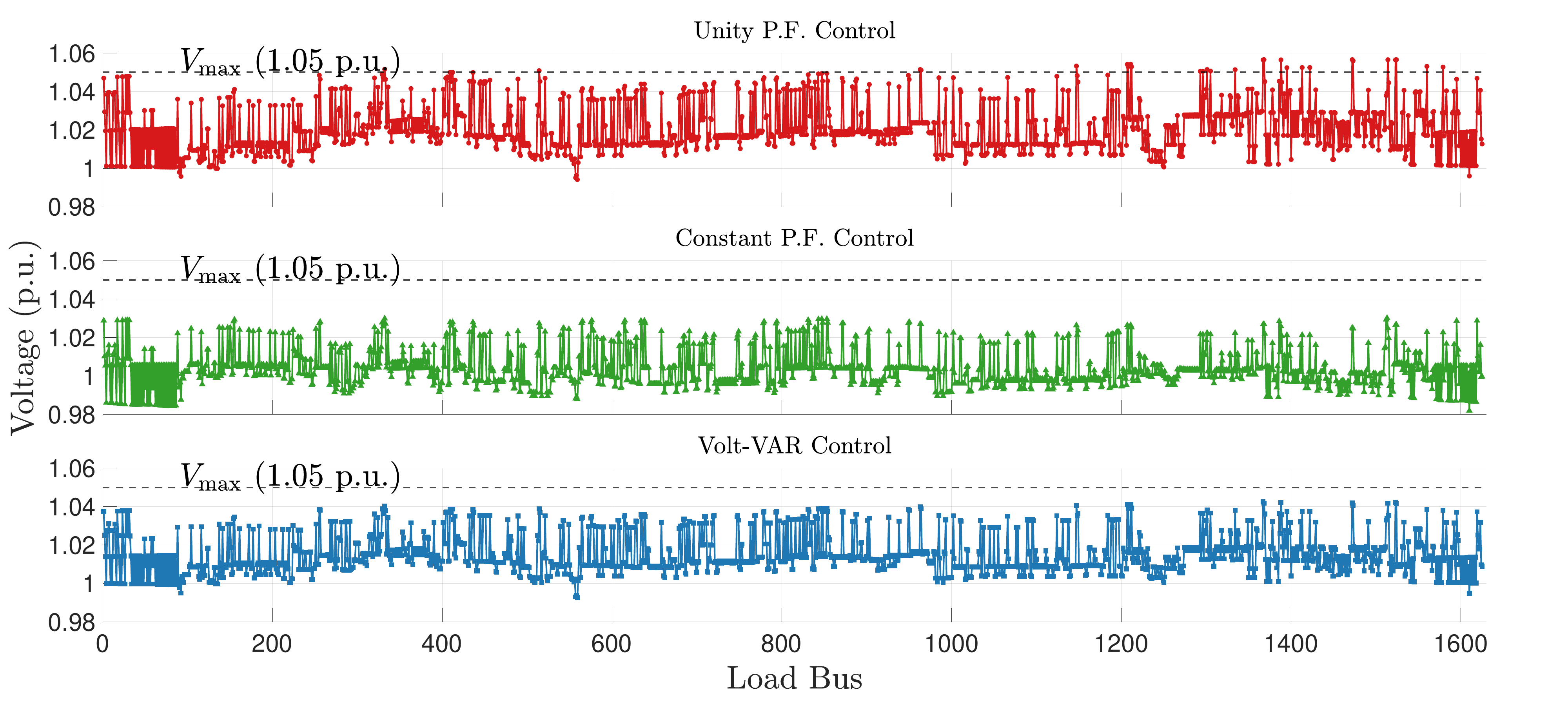}
    \caption{Steady-state voltage profiles across the feeder under three reactive control strategies (UPF, CPF, Volt-VAR) at a fixed active power injection per inverter.}
    \label{fig:voltage_comparison}
\end{figure*}

\subsection{Inverter-based Optimization Studies}
\label{subsec:generalizability}

\noindent 
%The results above demonstrate that the proposed \InverterShort{} model supports both detailed device-level analysis and scalable integration across large distribution networks. 
Next, we study the performance of \InverterShort{}  in two optimization settings: scheduling Home Energy Management System (HEMS) and grid-level PV export curtailment. We retain the objective and prosumer accounting from  \cite{badmus2024anoca}, but replace the inverter block as follows. 
(i) We substitute the AC-side charge/discharge pair $(P_{c,\tau},P_{d,\tau})$ with a single signed inverter AC power $P_\tau$ in the prosumer balance eqn. 
(ii) We update SOC using the inverter DC-port power together with the battery relation \eqref{eq:SOC_update}. 
(iii) We enforce the TSBI physics at each timestep using \eqref{eq:fsc_phys_sum}--\eqref{eq:lcl_phys_sum}. 

% Unless stated otherwise, we set $Q_\tau=0$ (UPF) through the PCC relation in \eqref{eq:pcc_link_sum}. 

\subsubsection{HEMS Multiperiod Optimization}
\label{subsec:hems_validation}

\noindent HEMS schedule problem solves a multi-period optimization to find setpoints for \InverterShort{}-tied photovoltaic (PV) generation $P_{pv}$ and battery storage to minimize net electricity costs.
\cite{badmus2024validation} documents the problem formulation used in this experiment. 
For different scenarios, we only replace the inverter model in that paper with three different inverter models for optimization.
Every other component and problem objective remains exact.

%All formulations solve the following cost minimization problem:
%\begin{subequations}
%\begin{alignat}{2}
%    &\min \sum_{t \in \mathcal{T}} \left( C_{i,t} P_{i,t} - C_{e,t} P_{e,t} \right) \\
%    &\text{s.t.} \quad P_{\text{load},t} + P_{e,t} = P_{i,t} + P_{\text{solar},t} + P_{\text{inv},t}, \quad \forall t \in \mathcal{T} \\
%    &0 \leq \text{SOC}_t \leq \overline{\text{SOC}}, \quad \text{SOC}_0 = \text{SOC}_T = \text{SOC}_{\text{set}} \\
%    &\text{SOC}_{t+1} = \text{SOC}_t - \Delta \tau \cdot \frac{P_{\text{batt},t}}{E_{\max}}, \quad \forall t \in \mathcal{T}
%\end{alignat}
%\end{subequations}

%\noindent Here, $P_{i,t}$ and $P_{e,t}$ are the grid import and export powers at time $t$, respectively. $P_{\text{load},t}$ is the household demand, and $P_{\text{solar},t}$ is the solar PV generation. $P_{\text{inv},t}$ is the net inverter power, while $P_{\text{batt},t}$ is the net battery power; both positive when discharging, negative when charging. $\text{SOC}_t$ is the battery state of charge, bounded by $\overline{\text{SOC}}$, and initialized/finalized at $\text{SOC}_{\text{set}}$. $\Delta \tau$ is the time step duration, $E_{\max}$ is the battery energy capacity, and $C_{i,t}$ and $C_{e,t}$ are the electricity import and export prices. $\mathcal{T}$ denotes the set of time-periods.

\begin{added}{Reviewer 1 Comment 6. No changes, just pointing this part to the reviewer. Also, First sentence addresses part of Reviewer 2 Comment 6}

The first model, which is the Constant-Efficiency Binary (CE-B) model, uses the inverter formulation from \cite{badmus2024anoca}, which uses fixed charge/discharge efficiencies $(\eta_c, \eta_d)$ and enforces mutual exclusivity between battery charging $P_{c,t}$ and discharging $P_{d,t}$ at any time $t$ with a binary variable $z_t \in \{0,1\}$:
\begin{subequations}
\begin{equation}
0 \leq P_{c,t} \leq z_t \cdot \overline{P}, \quad
0 \leq P_{d,t} \leq (1 - z_t) \cdot \overline{P}
\end{equation}

 The second model, which is the Constant-Efficiency Complementarity-Slackness (CE-CS) model, also uses fixed efficiencies for inverter charge and discharge cycles; however, it enforces mutual exclusivity between battery charging and discharging with a complementarity constraint \cite{elsaadany2023battery}, allowing the problem to be formulated as a nonlinear program (NLP):
\begin{equation}
P_{c,t} \cdot P_{d,t} = \epsilon
\end{equation}

%The model also uses fixed efficiency for charge and discharge cycles and
%enforces mutual exclusivity with continuous variables with complementary slackness condition. 
%We refer to this formulation as CE-CS.
%\begin{equation}
%P_{c,t} \cdot P_{d,t} = 0
%\end{equation}

\noindent CE-B and CE-CS compute the net inverter and battery power at time $t$ as:
\begin{equation}
P_{\text{inv},t} = P_{pv,t} + P_{batt,t}, \quad
P_{batt,t} = P_{d,t}/\eta_d - \eta_c P_{c,t}
\end{equation}
\end{subequations}

\noindent The final formulation uses the \InverterShort{} model from this paper.
It defines inverter power using a single continuous variable $P_{\text{inv},t}$, avoiding the need to separately model charge and discharge operations or impose mutual exclusivity constraints. 
To isolate the impact of active power control, we evaluate all models at unity power factor at each time step, i.e., $Q_{\text{inv},t} = 0$.
\end{added}

We run a 7-day, 5-minute resolution look-ahead optimization using load and solar data from a New York feeder \cite{almassalkhi2020hierarchical}, New York electricity tariffs from the IEA \cite{IEA2022Electricity}. 
% and assume a maximum continuous power of 5 kW for TSBI together with inverter rating to be 11.5KVA. 
We assume a maximum continuous charging and discharging power of 5kW and an apparent power rating of 11.5kVA to accommodate hybrid PV-battery operation.
We compute energy-weighted TSBI charging and discharging efficiencies to ensure a fair comparison. 
We apply them to both CE-B and CE-CS formulations, as they both assume fixed inverter efficiency. 
The CE-B problem was solved using Gurobi \cite{gurobi2021gurobi}, while CE-CS and TSBI were solved using IPOPT \cite{wachter2006implementation}.

\begin{table}[htbp]
\centering
\scriptsize
\setlength{\tabcolsep}{4.5pt}
\caption{Objective cost comparison across HEMS formulations (in \$) and average solve time.}
\label{tab:hems_objectives}
\begin{tabular}{@{}lcccccccc@{}}
\toprule
\textbf{Model} & \textbf{Day1} & \textbf{Day2} & \textbf{Day3} & \textbf{Day4} & \textbf{Day5} & \textbf{Day6} & \textbf{Day7} & \textbf{Avg. Time (s)} \\
\midrule
\textbf{CE-CS} & 9.20 & 16.14 & 9.82 & 11.17 & 15.03 & 16.02 & 20.58 & 1.09 \\
\textbf{CE-B}  & 7.72 & 13.25 & 8.26 & 8.60 & 13.19 & 14.67 & 19.05 & 0.20 \\
\textbf{TSBI}  & 7.83 & 13.40 & 8.36 & 8.75 & 13.30 & 14.85 & 19.18 & 36.40 \\
\bottomrule
\end{tabular}
\end{table}
\vspace{0.5em}

\begin{figure}[htpb]
    \centering
    \includegraphics[width=1\linewidth]{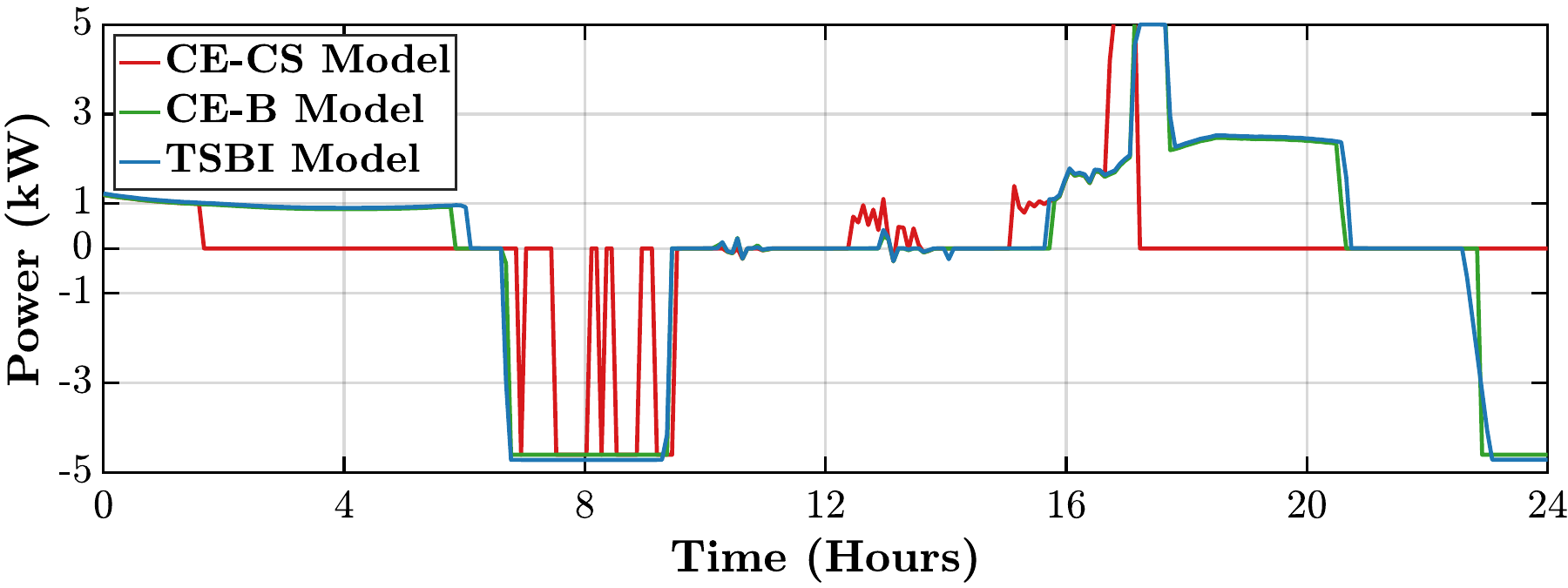}
    \caption{Comparison of battery dispatch profiles between the proposed TSBI and the constant-efficiency models (CE-B and CE-CS) for Day 5.}
    \label{fig:hems_dispatch}
    \vspace{0.5em}
\end{figure}

Table \ref{tab:hems_objectives} shows the objective value (net cost of electricity) for the same optimization problem (\cite{badmus2024anoca}) with different inverter models.
We observe CE-CS has the highest cost amongst the three formulations.
This is potentially due to convergence to local minima in many scenarios.
Fig. \ref{fig:hems_dispatch} validates this observation by plotting the battery dispatch for Day 5. 
CE-B converges to the globally optimal solution (0.0\% duality gap) across all test days. 
However, its result differs slightly from the TSBI formulation because it models efficiency as fixed and is likely more erroneous.
\begin{added}{Reviewer 1 Comment 5}
Although the TSBI formulation incurs a longer solve time than the simplified models, its average runtime of 36.4s remains well below the 5-minute HEMS scheduling interval. This ensures that each optimization can still be executed in real time while preserving full inverter physics and control integration.
Note that this result does not include the distribution feeder, which is where most of the optimization’s computational burden arises (see results in Section \ref{subsec:pv_curtailment}).
\end{added}

%CE-CS and TSBI are both nonconvex, but CE-CS enforces mutual exclusivity through a highly nonconvex complementarity constraint, while CE-B solves to 
%The higher cost in CE-CS suggests convergence to a poor local minimum, since its setup matches CE-B except for that constraint. 
%TSBI, though also nonconvex, consistently produced results close to CE-B, indicating better solution quality. 
%A key difference is that TSBI models time-varying efficiency directly, unlike the fixed energy-weighted value used in CE-B. 
%This may explain the slight deviation, though we cannot guarantee it due to TSBI’s nonconvexity. 
%Moreover, as shown in Section \ref{subsec:accuracy}, inverter efficiency drops sharply at low power, but CE-B applies the same high efficiency across all time steps, overstating performance during such periods.
%The average runtimes across the 7-day horizon for the CE-B, CE-CS, and TSBI formulations are 0.16s, 1.09s, and 36.11s, respectively.

% \begin{figure}[htpb]
%     \centering
%     \includesvg[width=1\linewidth]{images/battery_power_comparison_day5.svg}
%     \caption{Comparison of battery dispatch profiles between the proposed TSBI and the constant-efficiency models (CE-B and CE-CS) for Day 5.}
%     \label{fig:hems_dispatch}
%     \vspace{0.5em}
% \end{figure}

% \subsubsection{Grid-Level PV Curtailment}

% We applied the model to a 20,043-bus feeder in Vermont 
% under 100% PV penetration. Inverters operated at 75% of 
% their nameplate capacity to reserve headroom for reactive power.

% Under UPF, curtailment reached 9.6% at 8.6kW injection. 
% With Volt-VAR control, curtailment dropped to 1.1% 
% (\autoref{fig:curtailment}). The optimization enforced 
% IEEE 1547 voltage limits without additional control devices.

% \medskip
% \noindent\textbf{Summary:}
% The \InverterShort{} model replicates efficiency trends, 
% scales to large feeders, and integrates into nonlinear 
% and MILP formulations with solver stability and minimal overhead.

\subsubsection{Optimal PV Export Curtailment}
\label{subsec:pv_curtailment}
\begin{added}{Reviewer 1 Comment 6}
Next, we conduct a large-scale optimization that minimizes PV curtailment in a three-phase feeder, thereby avoiding violation of three-phase AC grid constraints.
The exact formulation is in \cite{badmus2024anoca}. 
% We only replace the fixed efficiency inverter model with \InverterShort{} for this study.
The only difference is that we replace the inverter unity power factor model in \cite{badmus2024anoca} with the physics-based \InverterShort{} model, and we include two control modes: unity power factor and Volt-VAR.
%Concretely, PV-bus injections and curtailment calculations is the same as as \cite{badmus2024anoca}, reactive power follows either Volt-VAR mode or Unity PF mode under the apparent power $S$ rating constraint, and the inverter operation is enforced by \eqref{eq:fsc_phys_sum}--\eqref{eq:pcc_phys_sum}. All AC network equations and limits remain as in \cite{badmus2024anoca}.
\end{added}

%In this formulation, inverter power export is modeled explicitly via physical TSBI constraints, including nonlinear I–V behavior and internal losses. 
The optimization is run on a 20,043-bus real-world utility feeder in Vermont, which has 3,088 load buses. We consider a photovoltaic (PV) system with the inverter model connected to all the load buses. 
We assume each inverter $n$ wants to export $P^*_{\text{e},n}$ (75\% of its rated apparent power) but may only export $P_{\text{e},n}$ due to voltage or line flow constraints. 
The optimization, therefore, minimizes total curtailed active power $(P_{cu}= P^*_{\text{e}} - P_{\text{e}})$ subject to 3-phase unbalanced AC power flow equality constraints and voltage and line flow inequality constraints. 
The objective has the following form:
\begin{equation}
\min \sum_{n \in \mathcal{N}_{\text{inv}}} \|P_{\text{cu},n}\|_{\infty} 
\label{eq:DMS_objective_summary}
\end{equation}

\noindent We use the $\|\cdot\|_{\infty}$ norm to ensure fair distribution of curtailment by minimizing the worst-case curtailed power across all inverters. We evaluate two reactive control strategies, unity power factor (UPF) and Volt-VAR. The Volt-VAR control is  based on IEEE 1547-2018 Category A settings  \cite{photovoltaics2018ieee}, and assesses system flexibility under three inverter rating levels. Under UPF, each inverter exports only active power. In contrast, Volt-VAR dynamically adjusts reactive power $Q_{\text{ctrl}}$ based on local voltage to support system voltage and reduce curtailment.

% \begin{figure}[htpb]
%     \centering
%     \includesvg[width=1\linewidth]{images/percentage_infeasibility_plot.svg}
%     \caption{Percentage of curtailed active power under Unity Power Factor (UPF) and Volt-VAR (Volt-VAR) control for different inverter power levels.}
%     \vspace{0.5em}
%     \label{fig:curtailment}
% \end{figure}
\begin{figure}[htpb]
    \centering
    \includegraphics[width=0.9\linewidth]{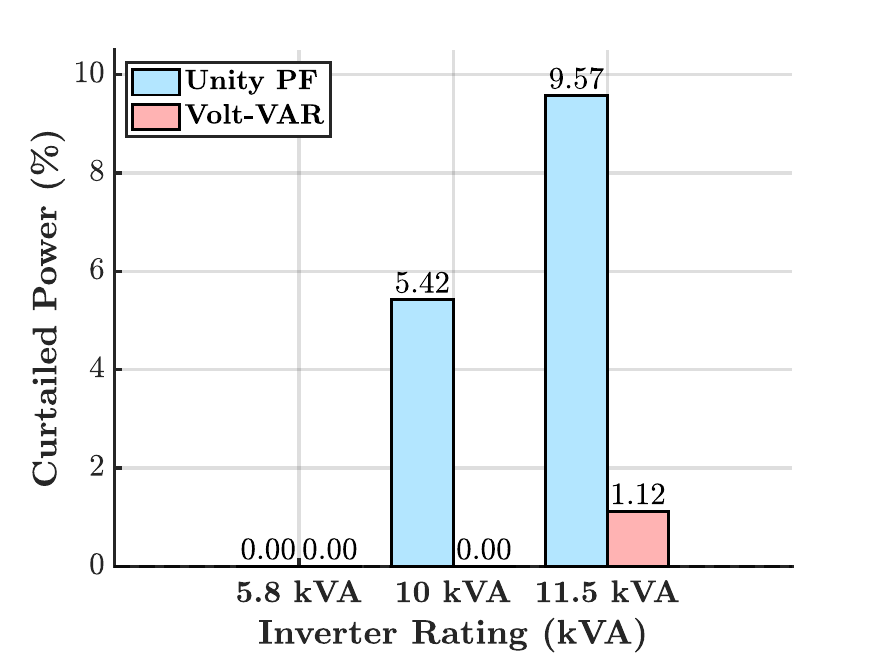}
    \caption{Percentage of curtailed active power under Unity Power Factor (UPF) and Volt-VAR control for different inverter power levels.}
    \vspace{0.5em}
    \label{fig:curtailment}
\end{figure}

 The UPF formulation includes 100,675 variables, 97,586 equality constraints, and 12,352 inequality constraints. For Volt-VAR, each of the 3,088 inverters introduces three additional equality constraints, increasing the equality constraint count and variable count by 3 times the number of inverters, yielding 106,850 equality constraints and 109,939 variables. 
 We embed all operational limits and the inverter's physical and control behavior directly in the physics-based formulation. Fig. \ref{fig:curtailment} shows that Volt-VAR consistently reduces curtailment across all inverter ratings. At 7.5 kVA (75\% of 10 kVA), UPF leads to 5.42\% curtailment, while Volt-VAR eliminates it. At 8.63 kW (75\% of 11.5 kVA), Volt-VAR limits curtailment to 1.12\%, compared to 9.57\% under UPF. These results confirm that voltage-dependent reactive control (implemented in \InverterShort{}) effectively reduces curtailment and enhances grid flexibility.
\begin{added}{Editor Comment 5}
In addition, across the three $P^{*}_{e,n}$ scenarios, the UPF and Volt–VAR formulations yield average solve times of 385s and 341s, respectively. The similar solve times indicate that incorporating voltage-dependent reactive control does not introduce a significant computational burden. These results confirm that the large-scale three-phase optimization with embedded \InverterShort{} models and control modes remains tractable.
\end{added}

\section{Conclusions}\label{Conclusions}
% \noindent This work presented a steady-state model of a two-stage bidirectional inverter (TSBI) that connects converter-level circuit behavior with system-level power flow analysis and optimization. The model addresses key limitations in prior approaches by explicitly capturing switching and conduction losses, modeling continuous bidirectional power flow, and embedding control behavior as smooth functions within the formulation. Through validation against time-domain simulation and deployment in residential and feeder-level case studies, we demonstrated that the TSBI model offers significantly improved accuracy over existing models and can be applied in analysis and optimization tasks. These properties make the \InverterShort{} model suitable for integration into analysis and optimization tools used in planning and real-time operational studies for inverter-rich distribution networks.

\noindent In this paper, we present a novel steady-state equivalent circuit model for \Inverter{} (\InverterShort{}) that implicitly integrates the converter's physics, internal losses, and control functions without loss of generality.
We validate the model accuracy against a higher-fidelity time-domain model and deploy it in large-scale grid-level optimizations. 
The results demonstrate the model's accuracy, scalability, and ability to include control generally across large networks. 
% We draw the following key conclusions:
% \begin{itemize}
%     \item The \InverterShort{} model captures inverter loss and bidirectional behavior implicitly based on foundational physics-based constraints
%     \item The model supports continuous bidirectional power flow and embeds control logic for constant power, Volt-VAR, and MPPT modes using controlled sources.
%     \item used the model in multiple load nodes to perform three-phase power flow analysis with inverters and optimization studies.
% \end{itemize}
% \noindent

% \noindent Overall, these properties make the \InverterShort{} model suitable for integration into analysis and optimization tools used in planning and real-time operational studies for inverter-rich distribution networks.

\noindent We draw the following key conclusions:
\begin{itemize}[leftmargin=1em]
    \item \textbf{Accuracy:} The model accurately captures the inverter losses for different operating conditions by capturing the semiconductor-level physics of the switches.
    \item \textbf{Modeling Generality:} The model supports any inverter control mode by embedding control logic into a pair of current-controlled sources as twice-differentiable continuous functions.
    \item \textbf{Scalability:} The model supports large-scale simulation and optimization as shown via a run on a 20k+ nodes Vermont network.
\end{itemize}

%\noindent These features make the \InverterShort{} model practical for integration into large-scale distribution grid planning and real-time operational tools, particularly in inverter-dense networks.

\section{Acknowledgement}
\noindent We acknowledge Muhammad Hamza Ali for his contributions to developing the equivalent-circuit-based distribution power-flow framework used in this work. We also acknowledge Peng Sang for technical discussions and contributions towards equivalent circuit modeling of PV and battery systems.

\bibliographystyle{IEEEtran_custom}
\bibliography{Refrences}

% \begin{IEEEbiographynophoto}{Emmanuel Badmus} is a Ph.D. student in Electrical Engineering at the University of Vermont. His research focuses on power flow and optimal power flow in distribution systems.
% \end{IEEEbiographynophoto}
% \begin{IEEEbiographynophoto}{Amritanshu Pandey} is an Assistant Professor at the University of Vermont. His research focuses on power system simulation, optimization, and energy equity.
% \end{IEEEbiographynophoto}

% \begin{IEEEbiography}[{\includegraphics[width=1in,height=1.25in,clip,keepaspectratio]{images/fig1.png}}]{IEEE Publications Technology Team}
% In this paragraph you can place your educational, professional background and research and other interests.\end{IEEEbiography}

\end{document}